\documentclass[aps,pra,12pt,nofootinbib,superscriptaddress,longbibliography]{revtex4-1}
\pdfoutput=1

\usepackage[utf8]{inputenc}
\usepackage{amssymb,amsmath,amsfonts,amsthm}
\usepackage{verbatim}
\usepackage{enumerate}
\usepackage{tensor}
\usepackage{fancyhdr}
\usepackage{graphicx}
\graphicspath{{pics/}{cnot-pics/}}

\DeclareMathOperator{\trace}{Tr}  

\usepackage{color}
\definecolor{dred}{rgb}{.8,0.2,.2}
\definecolor{ddred}{rgb}{.8,0.5,.5}
\definecolor{dblue}{rgb}{.2,0.2,.8}
 
\newcommand{\vb}[1]{}

\usepackage{hyperref}
\hypersetup{
    colorlinks,
    linkcolor={red!50!black},
    citecolor={blue!50!black},
    urlcolor={blue!80!black}
}

\theoremstyle{plain}
\newtheorem{theorem}{Theorem}

\theoremstyle{definition}
\newtheorem{definition}[theorem]{Definition}
\newtheorem{example}[theorem]{Example}

\newcommand{\bra}[1]{\mbox{$\langle #1|$}}
\newcommand{\ket}[1]{\mbox{$|#1\rangle$}}
\newcommand{\braket}[2]{\mbox{$\langle #1|#2\rangle$}}
\newcommand{\ketbra}[2]{\mbox{$|#1\rangle\langle #2|$}}
\usepackage[colorinlistoftodos]{todonotes}

\newcommand{\inprod}[2]{\ensuremath{\left\langle #1, #2 \right\rangle}}

\DeclareMathOperator{\Tr}{Tr}

\DeclareMathOperator{\rank}{rank}

\newcommand{\isom}{\cong}  

\newcommand{\x}{\mathbf{x}}
\newcommand{\y}{\mathbf{y}}

\newcommand{\I}{\openone}     
\newcommand{\R}{{\mathbb R}}  
\newcommand{\C}{{\mathbb C}}  
\newcommand{\K}{{\mathbb K}}  
\newcommand{\swap}{{\sf{SWAP}}}
\newcommand{\cnot}{{\sf{CNOT}}}
\newcommand{\had}{{\sf{H}}}
\newcommand{\XOR}{{\sf{XOR}}}
\newcommand{\COPY}{{\sf{COPY}}}

\newcommand{\spidx}[3]{^{(#1)}{}\indices{^{#2}_{#3}}}
\newcommand{\bv}{e}     
\newcommand{\dv}{\eta}  

\def\1#1{{\bf #1}}
\def\2#1{{\cal #1}}
\def\3#1{{\sl #1}}
\def\4#1{{\tt #1}}
\def\5#1{{\sf #1}}
\def\6#1{{\mathfrak #1}}
\def\7#1{{\mathbb #1}}

\usepackage{tikz}
\usepackage{lipsum}
\usetikzlibrary{shapes.geometric}

\newcommand{\be}{\begin{equation}}
\newcommand{\ee}{\end{equation}}

\newlength{\figtextvoffset}
\newcommand{\fscale}{1.0}

\newcommand{\fhelp}[1]{\scalebox{\fscale}{$#1$}}
\newcommand{\f}[1]{\settoheight{\figtextvoffset}{\fhelp{#1}}\raisebox{-0.5\figtextvoffset}{\fhelp{#1}}}

\usepackage[framemethod=default]{mdframed}
\mdfdefinestyle{examplestyle}{%
skipabove=3mm,skipbelow=3mm,%
linecolor=green,linewidth=3pt,%
frametitlerule=true,%
frametitlebackgroundcolor=gray!20,%
innertopmargin=2mm,%
splittopskip=2mm,%
repeatframetitle=false,
}

\hypersetup{
  pdfinfo={
    Title={Quantum Tensor Networks in a Nutshell},
    Author={J.D. Biamonte and V. Bergholm},
    Subject={tensor network tutorial},
    Keywords={tensor networks; matrix product states; string diagrams; quantum circuits; tree tensor networks; introduction; review; tutorial},
  }
}

\begin{document}

\newlength{\diagwidth}
\setlength{\diagwidth}{10cm }

\title{Quantum Tensor Networks in a Nutshell}

\author{Jacob Biamonte}
\email{jacob.biamonte@qubit.org}
\homepage{www.QuamPlexity.org}
\affiliation{Quantum Software Initiative\\ Skolkovo Institute of Science and Technology, Skoltech Building 3, Moscow 143026, Russia}
\affiliation{Institute for Quantum Computing \\ University of Waterloo, Waterloo, N2L 3G1 Ontario, Canada}

\author{Ville Bergholm}
\email{ville.bergholm@iki.fi}
\affiliation{Quantum Software Initiative\\ Skolkovo Institute of Science and Technology, Skoltech Building 3, Moscow 143026, Russia}


\begin{abstract}
{\bf
Tensor network methods are taking a central role in modern quantum physics and beyond. They can provide an efficient approximation to certain classes of quantum states,
and the associated graphical language makes it easy to describe and pictorially reason about quantum circuits, channels, protocols, open systems and more. Our goal is to explain tensor networks and some associated methods as quickly and as painlessly as possible. Beginning with the key definitions, the graphical tensor network language is presented through examples. We then provide an introduction to matrix product states.  
We conclude the tutorial with tensor contractions evaluating combinatorial counting problems.  The first one counts the number of solutions for Boolean formulae, whereas the second is Penrose's tensor contraction algorithm, returning the number of
$3$-edge-colorings of $3$-regular planar graphs.
}

\begin{center}
  \includegraphics[width=0.8\diagwidth]{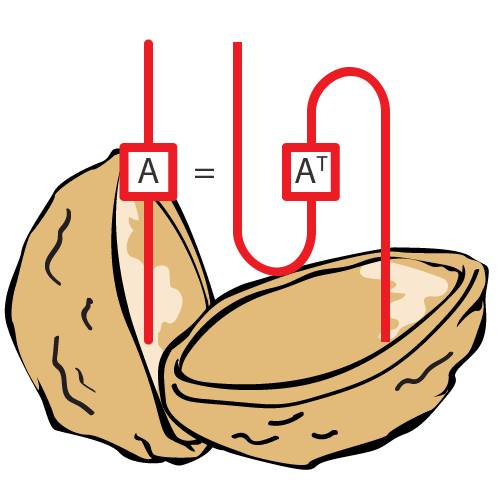}  
\end{center}
\end{abstract}

\maketitle

\section{Quantum Legos} 

Tensors are a mathematical concept that encapsulates and generalizes the idea of
multilinear maps, i.e.~functions of multiple parameters that are linear with respect to every parameter.
A tensor network is simply a countable collection of tensors connected by contractions.
`Tensor network methods' is the term given to the entire collection of associated tools, which are regularly employed in modern quantum information science, condensed matter physics, mathematics and computer science.  

Tensor networks come with an intuitive graphical language that can be used to reason about them.
This diagrammatic language
dates back to at least the early 1970s by Roger Penrose~\cite{Penrose}.
These methods have seen many advancements and adaptations to different domains of physics, mathematics and computer science.
An important milestone was David Deutsch's use of the diagrammatic notation in quantum computing,
developing the \emph{quantum circuit} (a.k.a.~quantum computational network)
model~\cite{Deutsch73}.
Quantum circuits are a special class of tensor networks, in which the arrangement of the tensors and their types are restricted.
A related diagrammatic language slightly before that is due to Richard Feynman~\cite{Feynman1986}.
The quantum circuit model---now well over two decades old---is widely used 
to describe quantum algorithms and their experimental implementations,
to quantify the resources they use (by e.g.~counting the quantum gates required),
to classify the entangling properties and computational power of specific gate families, and more.

There is now a lot of excitement about tensor network algorithms---for reviews
see~\cite{2014AnPhy.349..117O, Vidal2010, MPSreview08, TNSreview09, 2011AnPhy.326...96S, 2010arXiv1006.0675S, Schollw, 2014EPJB...87..280O,2013arXiv1308.3318E,2011JSP...145..891E, 2016arXiv160303039B, MAL-059, MAL-067}.
Some of the best known applications of tensor networks are 1D Matrix Product States (MPS),  Tensor Trains (TT), Tree Tensor Networks (TTN),
the Multi-scale Entanglement Renormalization Ansatz (MERA), Projected Entangled Pair States (PEPS)---which generalize matrix product states to higher dimensions---and various other renormalization methods~\cite{Vidal2010, MPSreview08, TNSreview09, 2011AnPhy.326...96S, 2009PhRvL.102e7202H, 2013arXiv1308.3318E, MAL-059}.
The excitement is based on the fact that certain
classes of quantum systems can now be simulated more efficiently, studied in greater detail,
and this has opened new avenues for a greater understanding of certain physical systems.

These methods approximate a complicated quantum state using a tensor network with a simplistic, regular structure---essentially applying lossy data compression that preserves the most important properties of the quantum state.
To give the reader a rough idea how these methods work, below we conceptually depict how the quantum state~$\psi$
could be represented (or approximated) using tensor networks in various ways.
\be
 \includegraphics{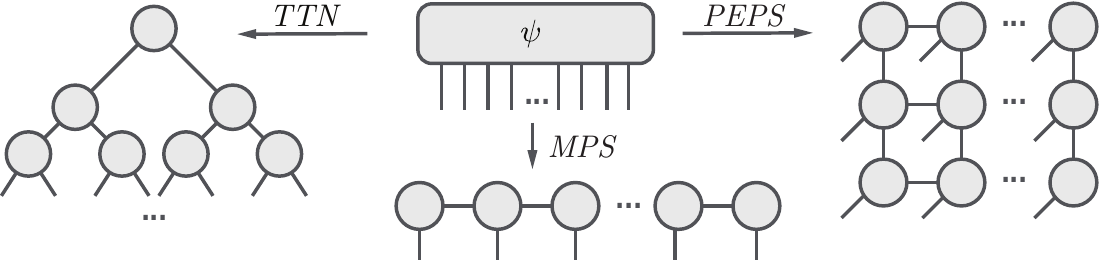}
\ee
We assume that most readers will have a basic understanding of some quantum theory, linear algebra and tensors.
In Appendix~\ref{app:tensor}, we provide a short mathematical definition of tensors and tensor products.
However, readers may wish to skip these definitions and for now proceed with a more informal or intuitive
understanding of the idea.

There are several notational approaches to tensors.
We begin by using abstract index notation, and then explain how it connects to the Dirac notation used in quantum computing.
The connection between tensor networks and quantum circuits is elucidated at the end of Section~\ref{sec:penrose}. 

\tableofcontents


\section[Tensors to Networks]{From Tensors to Networks} \label{sec:penrose}

\paragraph{{\bf Drawing tensors.}}
In the tensor diagram notation,
a tensor is a labelled shape such as a box, oval or triangle,
with zero or more open output legs (or \emph{arms}) pointing up, and zero or more
open input legs pointing down.
Individual arms and legs each correspond to upper and lower indices, respectively.\footnote{
This conformity is simplest to get started.
However, to conserve space, tensor diagrams are often rotated without
any warning $90$~degrees clockwise. In practice this should
be clear from the context.}
The arm and leg wires may be labelled with the indices they represent,
or with the vector spaces they correspond to, if necessary.
An order-$(0,0)$ tensor without any open arms or legs is
simply a complex number.
\be
 \includegraphics{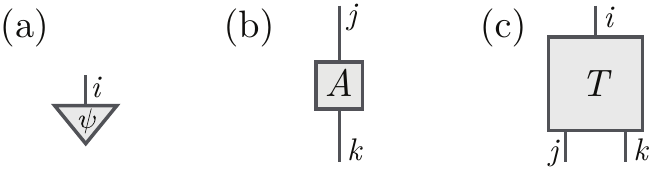} 
\ee
For example, diagram (a) above represents the tensor~$\psi^i$ with a
single upper index (a vector),
diagram (b) the tensor~$A\indices{^{j}_{k}}$ (a matrix),
and diagram (c) the tensor~$T\indices{^i_{jk}}$.

\paragraph{{\bf Tensor juxtaposition.}}

When two or more disconnected tensors appear in the same diagram they are multiplied
together using the tensor product.
In quantum physics notation, they would have a tensor
product sign~$\otimes$ between them.
In the abstract index notation the tensor product sign is omitted.

Tensors
can be freely moved past each other.  This is sometimes called planar deformation.
\be
 \includegraphics{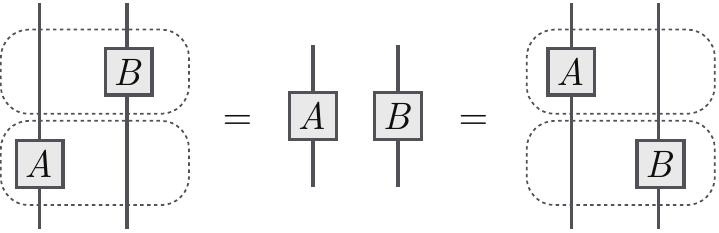}
\ee
From the diagram above, using equations we have 
\begin{equation}
 (\I\otimes B)(A\otimes \I) = A\otimes B = (A\otimes \I)(\I\otimes B),
\end{equation}
where we make use of the wire also playing the role of the identity tensor~$\I$---detailed in Section~\ref{sec:wires}.
As we shall soon see, wires
are allowed to cross tensor symbols and other wires,
as long as the wire endpoints are not changed.
This is one reason why tensor diagrams are often
simpler to deal with
than their algebraic counterparts.
\be
 \includegraphics{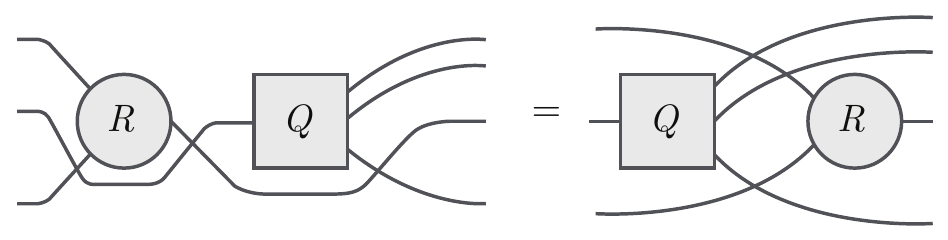}
\ee
In the diagram above we did not label the wires, since
it is an arbitrary assignment. If we did, we could for example denote it as
$Q\indices{^{deg}_{b}}R\indices{^{f}_{ac}}$.

\paragraph{{\bf Connecting wires.}}


Connecting two tensor legs with a wire means that the
corresponding indices are contracted (summed over).
\be
 \includegraphics{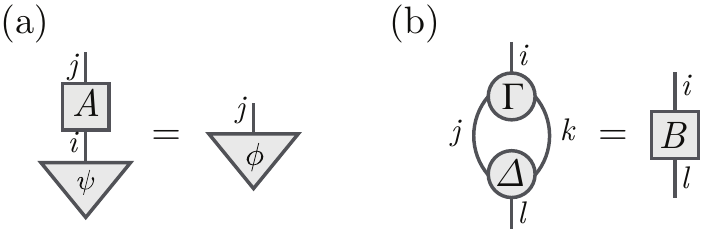}
\ee
In diagram (a) above  we find a matrix multiplying a vector
which results in another vector.
Diagram (a) is equivalent to the expression
\begin{equation}
  A\indices{^j_i} \psi^i = \phi^j,
\end{equation}
where we notice that the wire labeled~$i$ is fully connected, and
hence the corresponding index is summed over.
We used the \emph{Einstein summation convention},
in which any index that appears exactly twice in a term is summed over.

In (b) we face a slightly more complicated
case, departing from familiar vectors and matrices, contracting two
indices between two order-3 tensors. Diagram (b)
is equivalent to
\begin{equation}
 \Gamma\indices{^i_{jk}} \Delta\indices{^{jk}_{l}} = B\indices{^i_l}.
\end{equation}
Together, two or more tensors in a diagram form a \emph{tensor network}.
If none of the tensors have any open arms or legs
the network is said to be fully contracted: it evaluates to some
complex number, a scalar.

You have probably noticed that we are being strict and are leaving spaces to
identify the order of indices as we read them left to right across the
page (e.g.~in $\Gamma\indices{^{i}_{jk}}$ we have $i$, followed by $jk$).
In Section~\ref{sec:wires} we are going to talk about this in more detail,
and justify why it is not necessary: we are doing it here for ease of
illustration.

Some authors like to put arrows on the tensor diagram wires~\cite{2009arXiv0908.2469B},
to denote the difference between vector spaces and their duals.\footnote{See Appendix~\ref{app:tensor} for the definition of a dual space.}
In this work we mostly deal with finite-dimensional vector spaces over real or complex numbers,
and for each vector space~$V$ we pick a preferred basis (called the \emph{computational basis} in quantum computing).
The basis can be used to define an inner product,
turn~$V$ into a Hilbert space,
and establish a bijective mapping between~$V$ and its dual~$V^*$.
We then use this mapping to equate dual vectors in~$V^*$ with their counterparts in~$V$.
In this case arrows on the wires add nothing essential, so we omit them.

\paragraph{{\bf Connection to quantum computing notation.}}
As mentioned, tensors are multilinear maps.
They can be expanded in any given basis, and expressed in
terms of their components.
In quantum information science one often introduces a \emph{computational basis}
$\{\ket{k}\}_k$
for each Hilbert space
and expands the tensors in it, using kets ($\ket{~}$) for vectors and bras ($\bra{~}$) for dual vectors:
\be 
T = \sum_{ijk}T\indices{^i_{jk}}\ket{i}\bra{jk}.
\ee 
Here $T\indices{^i_{jk}}$ is understood
not as abstract index notation but as the actual components of the
tensor in the computational basis.
In practice there is little room for confusion.
The Einstein summation convention is rarely used in quantum information science,
hence we write the sum sign explicitly.

So far we have explained how tensors are represented in tensor diagrams,
and what happens when wires are connected.
The ideas are concluded by four
examples; we urge the reader to work through the examples and check the results for themselves.

The first example introduces a familiar structure from linear algebra in tensor form.
The next two examples come from quantum entanglement theory---see connecting
tensor networks with invariants~\cite{2013JPhA...46U5301B, 2014SIGMA..10..095C}.
The fourth one showcases quantum circuits, a subclass of tensor networks
widely used in the field of quantum information.
The examples are chosen to illustrate properties of tensor networks and should be self-contained. 

\begin{example}[The $\epsilon$ tensor]\label{ex:epsilon}
A tensor is said to be fully antisymmetric if swapping any pair of
indices will change its sign: $A_{ij} = -A_{ji}$.
The $\epsilon$ tensor is used to represent the fully antisymmetric Levi-Civita
symbol, which in two dimensions can be expressed as
\begin{equation}
 \epsilon_{00}=\epsilon_{11}=0, \qquad
 \epsilon_{01}=-\epsilon_{10}=1.
\end{equation}
The $\epsilon$ tensor can be used to compute the determinant of a matrix. In two dimensions we have
\be
\det(S) = \epsilon_{ij} S\indices{^i_0} S\indices{^j_1}.
\ee
Using this we obtain
\be
 \includegraphics{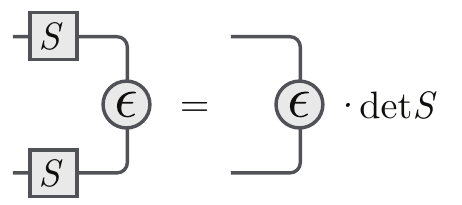}
\ee
as can be seen by labeling the wires in the diagram.
In equational form this is
\begin{equation}
\epsilon_{ij} S\indices{^i_m} S\indices{^j_n} = \det(S) \: \epsilon_{mn}.
\end{equation}
In terms of quantum mechanics, $\epsilon$ corresponds to the two-qubit
singlet state:
\be\label{eqn:singlet}
\frac{1}{\sqrt{2}} \ket{\epsilon} = \frac{1}{\sqrt{2}}(\ket{01}-\ket{10}).
\ee
This quantum state is invariant under any transformation of the form $U \otimes U$, where $U$~is a $2\times 2$ unitary,
as it only gains an unphysical global phase factor~$\det(U)$.

\end{example}

\begin{example}[Concurrence and entanglement]\label{ex:concurrence}
Given a two-qubit pure quantum state $\ket{\psi}$,
its \emph{concurrence}
$C(\psi) = |C'(\psi)|$
is the absolute value of the following tensor network expression~\cite{concurrence}:
\be\label{fig:con}
 \includegraphics[width=0.36\textwidth]{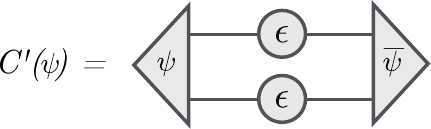}
\ee
Here $\overline{\psi}$ is the complex conjugate of~$\psi$ in the computational basis.
The concurrence is an entanglement monotone, a function from states to
nonnegative real numbers that measures how entangled the state is.
$\ket{\psi}$ is entangled if and only if
the concurrence is greater than zero.

Consider now what happens when we act on $\ket{\psi}$ by an arbitrary local unitary operation, i.e.~$\ket{\psi'} = (U_1 \otimes U_2) \ket{\psi}$.
\vb{We haven't yet shown how wires are bent!}
Using the result of Example~\ref{ex:epsilon} we obtain
\be
C\left((U_1\otimes U_2) \ket{\psi}\right) = C(\psi) |\det(U_1) \det(U_2)|.
\ee
Due to the unitarity $|\det U_1| = |\det U_2| = 1$, which means that the value of the concurrence
is \emph{invariant} (i.e.~does not change) under local unitary transformations.
This is to be expected, as local unitaries cannot change the amount of entanglement in a quantum state.
We will revisit concurrence in Example~\ref{ex:con-2}.

More complicated invariants can also be expressed as tensor networks~\cite{2013JPhA...46U5301B}.
We will leave it to the reader to write the following network as an algebraic expression:
\be
 \includegraphics[width=0.6\textwidth]{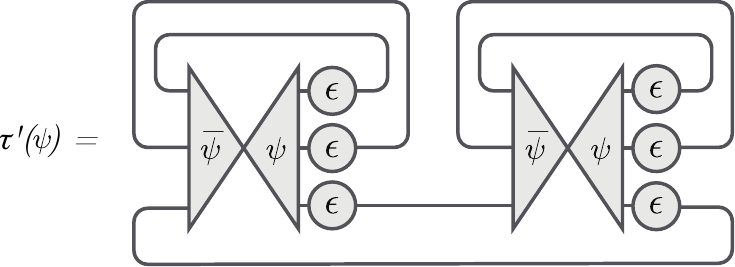}
\ee
If $\ket{\psi}$ is a 3-qubit quantum state,
$\tau(\psi) = 2 |\tau'(\psi)|$ represents the entanglement invariant known as the 3-tangle~\cite{2000PhRvA..61e2306C}.
It is possible to form invariants also without using the epsilon tensor.  For example,
the following expression represents the 3-qubit entanglement invariant known as the Kempe invariant~\cite{kempe}:
\begin{equation}\label{eqn:kempe}
K(\psi) = \psi^{ijk}~\overline{\psi}_{ilm}~\psi^{nlo}~\overline{\psi}_{pjo}~\psi^{pqm}~\overline{\psi}_{nqk}.
\end{equation}
The studious reader would draw the equivalent tensor network.
\end{example}

\begin{example}[Quantum circuits] \label{ex:circuits-1}

Quantum circuits are a restricted subclass of tensor networks that is widely used in the field of quantum information.
In a quantum circuit diagram each horizontal wire represents the Hilbert space associated with a quantum subsystem,
typically a single qubit.
The tensors attached to the wires represent unitary propagators acting on those subsystems, and are called \emph{quantum gates}.
Additional symbols may be used to denote measurements.
The standard notation is described in~\cite{PhysRevA.52.3457}.

Here we will consider a simple quantum circuit that can generate entangled Bell states.
It consists of two tensors, a Hadamard gate
(\had{})
and a controlled NOT gate
(\cnot, denoted by the symbol inside the dashed region):
\be
\label{eq:bell-circuit}
 \includegraphics{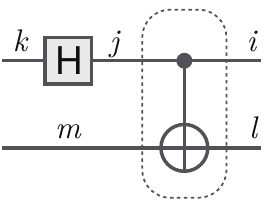}
\ee
The \cnot{} and Hadamard gates are defined as
\begin{align}
\cnot &= \sum_{ab} \ketbra{a, a \oplus b}{a,b} \quad \text{and}\\
\label{eqn:hadamard}
\had &= \frac{1}{\sqrt{2}} \sum_{ab} (-1)^{ab} \ketbra{a}{b},
\end{align}
where the addition in the \cnot{} is modulo~2.\footnote{Addition modulo 2: $1\oplus 1 = 0\oplus 0 = 0$, \; $1\oplus 0 = 0\oplus 1 = 1$.}
The reader should verify that acting on the quantum state~$\ket{00}$ the above circuit yields the Bell state
$\frac{1}{\sqrt{2}}(\ket{00}+\ket{11})$, and acting on~$\ket{11}$ it yields the singlet state
$\frac{1}{\sqrt{2}}(\ket{01}-\ket{10})$.
\end{example}

\begin{example} [\COPY{} and \XOR{} tensors] \label{ex:circuits-2}
One can view the \cnot{} gate itself as a contraction of two order-three tensors \cite{CTNS}:
\be 
\includegraphics{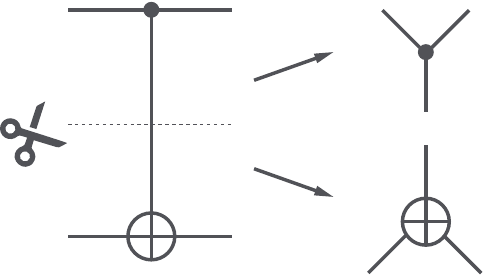}
\ee 
The top tensor ($\bullet$ with three legs) is called the \COPY{} tensor.
It equals unity when all the indices are assigned the same value ($0$ or $1$), and vanishes otherwise:
\be
\includegraphics{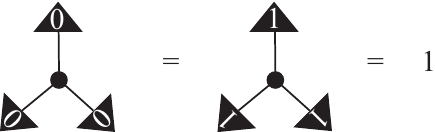}
\ee
Hence, \COPY{} acts to copy the binary inputs $0$ and~$1$:
\be
\includegraphics{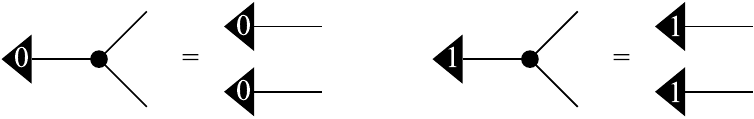}
\ee
The bottom tensor ($\oplus$ with three legs) is called the parity or \XOR{} tensor.
It equals unity when the index assignment contains an even number of $1$s, and vanishes otherwise:
\be
\includegraphics{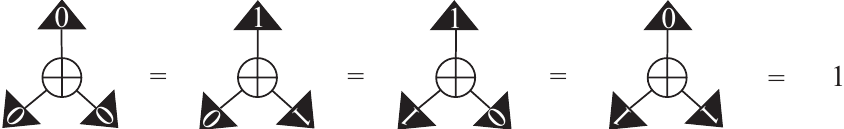}
\ee
The \XOR{} and \COPY{} tensors are related via the Hadamard gate as
\be
\label{eq:copy-vs-xor}
\renewcommand{\fscale}{1} \def\svgwidth{0.3\textwidth} 
\begingroup%
  \makeatletter%
  \providecommand\color[2][]{%
    \errmessage{(Inkscape) Color is used for the text in Inkscape, but the package 'color.sty' is not loaded}%
    \renewcommand\color[2][]{}%
  }%
  \providecommand\transparent[1]{%
    \errmessage{(Inkscape) Transparency is used (non-zero) for the text in Inkscape, but the package 'transparent.sty' is not loaded}%
    \renewcommand\transparent[1]{}%
  }%
  \providecommand\rotatebox[2]{#2}%
  \ifx\svgwidth\undefined%
    \setlength{\unitlength}{131.87398021bp}%
    \ifx\svgscale\undefined%
      \relax%
    \else%
      \setlength{\unitlength}{\unitlength * \real{\svgscale}}%
    \fi%
  \else%
    \setlength{\unitlength}{\svgwidth}%
  \fi%
  \global\let\svgwidth\undefined%
  \global\let\svgscale\undefined%
  \makeatother%
  \begin{picture}(1,0.35260936)%
    \put(0,0){\includegraphics[width=\unitlength]{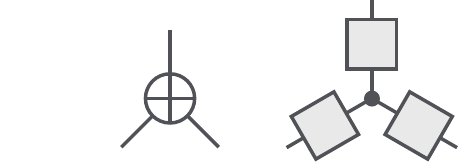}}%
    \put(0.52183271,0.11618927){\makebox(0,0)[lb]{\smash{=}}}%
    \put(0.23164872,0.14843194){\color[rgb]{0,0,0}\makebox(0,0)[b]{\smash{$\f{\frac{1}{\sqrt{2}}}$}}}%
    \put(0.81201669,0.25590749){\color[rgb]{0,0,0}\rotatebox{90}{\makebox(0,0)[b]{\smash{$\f{\sf{H}}$}}}}%
    \put(0.70963247,0.07857283){\color[rgb]{0,0,0}\rotatebox{-149.99999989}{\makebox(0,0)[b]{\smash{$\f{\sf{H}}$}}}}%
    \put(0.91440093,0.07857282){\color[rgb]{0,0,0}\rotatebox{-30.00000011}{\makebox(0,0)[b]{\smash{$\f{\sf{H}}$}}}}%
  \end{picture}%
\endgroup%
\quad
\ee
Thus one can think of \XOR{} as being a (scaled) copy operation in another basis:
\begin{subequations}
\begin{align}
        \frac{1}{\sqrt{2}}\XOR{} \ket{+} &= \ket{+}\ket{+},\label{eqn:xcopy}\\
        \frac{1}{\sqrt{2}}\XOR{} \ket{-} &= \ket{-}\ket{-},\label{eqn:xcopy2}
\end{align}
\end{subequations}
where $\ket{+} := \had\ket{0}$ and $\ket{-} := \had\ket{1}$.
In terms of components,
\begin{subequations}
\begin{align}
\COPY{}\indices{^{ij}_k}
&= (1-i)(1-j)(1-k)+ijk,\\
\XOR{}\indices{^{qr}_s} &= 1-(q+r+s)+2(qr+qs+sr)-4qrs.
\end{align}
\end{subequations}
The CNOT gate is now obtained as the tensor contraction
\begin{equation}\label{eqn:cnot}
\sum_m \COPY{}\indices{^{qm}_{i}} \: \XOR{}\indices{^r_{mj}} = \cnot{}\indices{^{qr}_{ij}}.
\end{equation}
The \COPY{} and \XOR{} tensors will be explored further in later examples and have many convenient properties~\cite{BB11,2012JPhA...45a5309D,B17}.

\end{example}

\section{Bending and Crossing Wires}\label{sec:wires}

\noindent ``\emph{It now ceases to be important to maintain a distinction 
between upper and lower indices.}''

{\hfill -- Roger Penrose, 1971 \cite{Penrose}}  \\

\paragraph{{\bf Cups and caps.}}
As explained in the previous section, wires are used to denote the contraction of pairs of tensor indices.
However, it is often useful to interpret certain wire structures as independent tensors of their own.
We start with three of these special \emph{wire tensors} that
allow one to rearrange the arms and legs of another tensor:
\be
\includegraphics{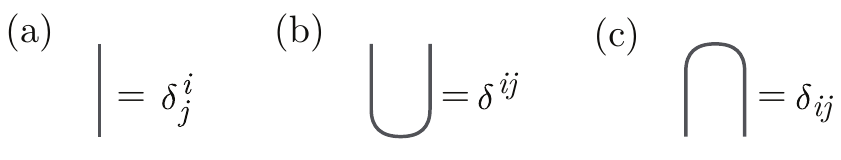}
\ee
The identity tensor (a) is used for index contraction
by connecting the corresponding legs. The cup (b) and the cap (c) raise and lower tensor indices
by bending the corresponding tensor legs.\footnote{Some readers familiar with relativity will note similarities with the metric tensor---here we will always work in a flat Euclidean space, meaning the metric tensors are trivial.}
Expanding them in the computational basis we obtain
\begin{align}
\I &= \sum_{ij} \delta\indices{^i_j}\ket{i}\bra{j} =\sum_k\ket{k}\bra{k},\\
 \ket{\cup} &= \sum_{ij} \delta\indices{^{ij}}\ket{ij} = \sum_k\ket{kk},\\
\bra{\cap} &= \sum_{ij} \delta\indices{_{ij}}\bra{ij} = \sum_k\bra{kk}.
\end{align}
In a quantum information context, the cup
also corresponds to an (unnormalized) Bell state, generalized so that it is not defined just for qubits.

\paragraph{{\bf Snake equation.}}\label{para:snake}
One can raise and then lower an index or vice versa, which
amounts to doing nothing at all. This idea
is captured diagrammatically by the so called \emph{snake} or
\emph{zig-zag equation}~\cite{Penrose}.
\be
 \includegraphics{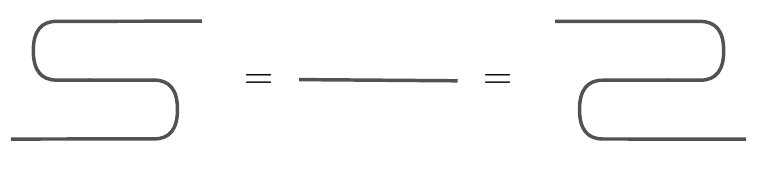}\label{eqn:snake}
\ee
In abstract index notation it is expressed succinctly as
$\delta^{ij} \delta_{jk} = \delta\indices{^i_{k}} = \delta_{kj} \delta^{ji}$.

\paragraph{{\bf $\swap$~gate.}}
Crossing two wires (as in diagram (a) below) can be thought of as swapping the relative order
of two vector spaces.
It corresponds to the $\swap$ gate used in quantum computing.
If both wires represent the same vector space, it can be alternatively
understood as swapping the states of the two subsystems.
\be\label{fig:sym}
 \includegraphics{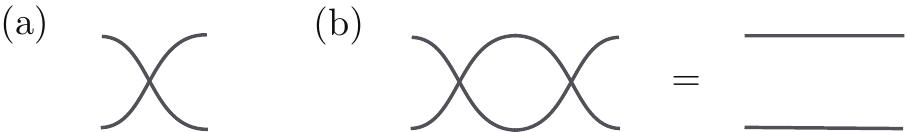}
\ee
Equation (b) illustrates that the $\swap$ operation is self inverse.
It may be written as
$\swap\indices{^{ij}_{kl}}=\delta\indices{^j_{k}}\delta\indices{^i_{l}}$,
or expanded in the computational basis as
$\swap = \sum_{ij} \ketbra{ij}{ji}$.
It also has a well-known implementation in terms of three CNOT gates as
\be
\includegraphics{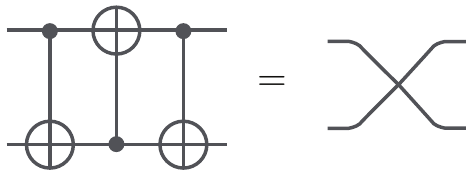}
\ee

\swap{} is the simplest nontrivial example of a permutation tensor. More
complicated permutations may be built out of the
$\delta\indices{^i_{j}}$ tensors in an obvious way.
We will return to this idea in Example~\ref{ex:spinor}.

\paragraph{{\bf Transpose.}} 
Given $A\indices{^i_j}$, we may reverse the positions of its indices using a cup and a cap.
This is equivalent to transposing the corresponding linear map in the computational basis:
\be
\label{eq:transpose}
\includegraphics{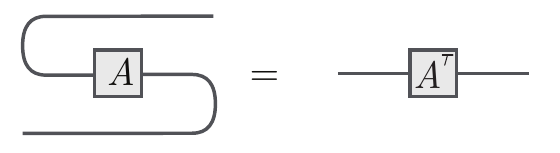}
\ee

\paragraph{{\bf Trace.}} 
In the tensor diagram notation, trace is given by appropriately joining
all the output wires of a tensor to corresponding input wires.
Diagram (a) below represents the trace~$A\indices{^i_i}$.
Diagram (b) represents the trace~$B\indices{^{iq}_{iq}}$.
\be
\includegraphics{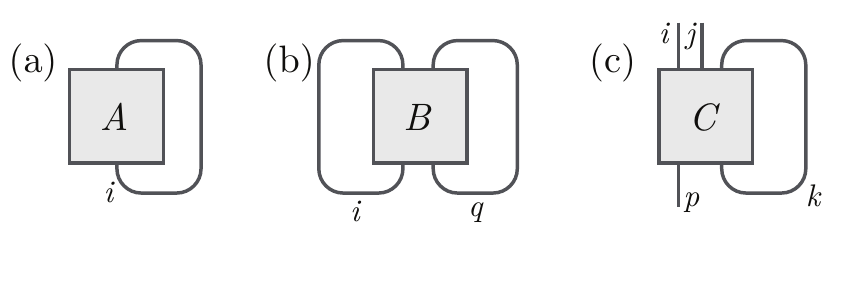}
\ee
Partial trace means contracting only some of the outputs with their
corresponding inputs, such as with the tensor $C\indices{^{ijk}_{pk}}$ shown in diagram (c).

\begin{example}[Partial trace]\label{ex:partial-trace}
The following is an early rewrite representing entangled pairs due to Penrose \cite{collected}.
\be
\includegraphics{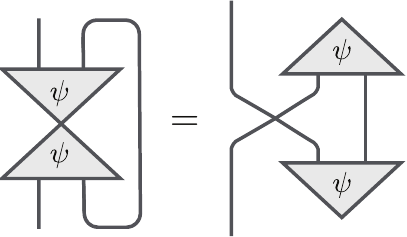}
\ee
The diagram on the left represents the partial trace of $\ketbra{\psi}{\psi}$ over the second subsystem.
Readers can prove that this equality follows by interpreting the bent wires as cups and caps, and the crossing wires as \swap{}s.
\end{example} 

\begin{example}[Partial trace of Bell states] Continuing on from Example \ref{ex:partial-trace}, if we choose $\ket{\psi} = \ket{\cup}$, i.e.~$\ket{\psi}$ is an unnormalized Bell state, we obtain 
\be
\includegraphics{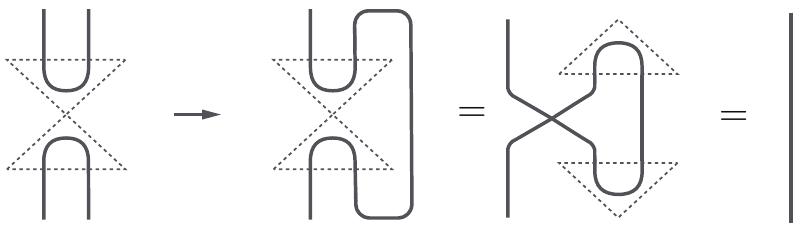}
\ee

\end{example}

\begin{example}[Relation between $\epsilon$ and \swap]\label{ex:spinor}


For any order-$2$ tensor $T^{ij}$ we can define its \emph{antisymmetrization}
as $T^{[ij]} = \frac{1}{2}(T^{ij}-T^{ji})$.
Here we used the notation of putting brackets around a group of indices---$[ij]$---to denote their antisymmetrization. 
Only indices of the same dimension may be antisymmetrized
(otherwise the expression would be undefined for some index values).

The fully antisymmetric $\epsilon$ tensor from Example~\ref{ex:epsilon}
has an interesting relation to the SWAP gate:
\be\label{eqn:spinor}
 \includegraphics{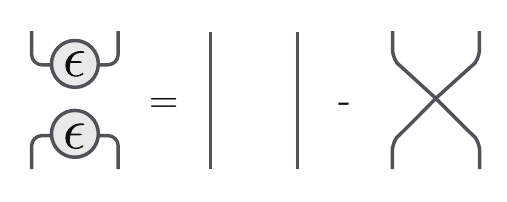}
\ee
or alternatively
\begin{equation}
\epsilon^{kl} \epsilon_{ij} =
\delta\indices{^k_i} \delta\indices{^l_j} -\delta\indices{^k_j} \delta\indices{^l_i}.
\end{equation}
It is now easy to show that for any tensor $T^{ij}$
(for which both indices are two-dimensional)
we can write
$T^{[kl]} = \frac{1}{2}\epsilon^{kl} \epsilon_{ij} T^{ij}$.

Both the concept of antisymmetrization and the epsilon tensor can be extended to more than two indices.
The \emph{antisymmetrizer} of $n$ $d$-dimensional vector spaces
is an order-$(n,n)$ tensor~$A\indices{^{i_1\cdots i_n}_{j_1 \cdots j_n}}$.
It can be expressed as the sum of all $n$-element permutations multiplied by their signatures.\footnote{The signature of a permutation $\sigma$ is $(-1)^{N(\sigma)}$, where $N(\sigma)$ is the number of pairwise swaps it contains.}
It antisymmetrizes $n$ $d$-dimensional indices by contraction:
\be
T^{[i_1\cdots i_n]} = A\indices{^{i_1\cdots i_n}_{j_1 \cdots j_n}} T^{j_1 \cdots j_n}.
\ee
When $d<n$ the only possible antisymmetric combination is a zero tensor,
and the corresponding $A$~vanishes identically.

For the general order-$(0,n)$ epsilon tensor, all the $n$ vector spaces need to be $n$-dimensional.
We then define
$\epsilon_{012 \ldots (n-1)} = 1$,
and all the other components are fixed by requiring complete antisymmetry,
i.e.~change of sign under the interchange of any two indices.
In particular, if any index value is repeated the corresponding component is zero.
Now
$\frac{1}{n!} \epsilon^{i_1\cdots i_n} \epsilon_{j_1 \cdots j_n}$
is an antisymmetrizer.

We shall use order-three epsilon tensors to count graph edge colorings by tensor contraction in Section~\ref{sec:color}.
\end{example}

\begin{example}[Quantum circuits for cups and epsilon states]\label{ex:state-prep}
The quantum circuit from Example~\ref{ex:circuits-1} is typically used to generate entangled qubit pairs.  
For instance, acting on the state~$\ket{00}$ yields the familiar Bell state---as a tensor network, this is equal to a normalized cup.
Here we also show the mathematical relationship the \XOR{} and \COPY{} tensors have with the cup
(here $\ket{+}:=\ket{0}+\ket{1}$):
\be \label{cnot-bell-states}
\renewcommand{\fscale}{1} \def\svgwidth{0.5\textwidth} 
\begingroup%
  \makeatletter%
  \providecommand\color[2][]{%
    \errmessage{(Inkscape) Color is used for the text in Inkscape, but the package 'color.sty' is not loaded}%
    \renewcommand\color[2][]{}%
  }%
  \providecommand\transparent[1]{%
    \errmessage{(Inkscape) Transparency is used (non-zero) for the text in Inkscape, but the package 'transparent.sty' is not loaded}%
    \renewcommand\transparent[1]{}%
  }%
  \providecommand\rotatebox[2]{#2}%
  \ifx\svgwidth\undefined%
    \setlength{\unitlength}{225.03125bp}%
    \ifx\svgscale\undefined%
      \relax%
    \else%
      \setlength{\unitlength}{\unitlength * \real{\svgscale}}%
    \fi%
  \else%
    \setlength{\unitlength}{\svgwidth}%
  \fi%
  \global\let\svgwidth\undefined%
  \global\let\svgscale\undefined%
  \makeatother%
  \begin{picture}(1,0.49618109)%
    \put(0,0){\includegraphics[width=\unitlength]{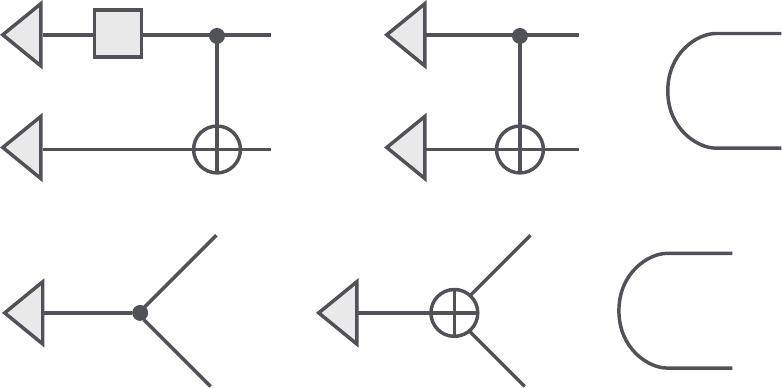}}%
    \put(0.01970215,0.29315906){\makebox(0,0)[lb]{\smash{0}}}%
    \put(0.42406555,0.08146269){\makebox(0,0)[lb]{\smash{0}}}%
    \put(0.01970215,0.43737859){\makebox(0,0)[lb]{\smash{0}}}%
    \put(0.69612554,0.08429385){\makebox(0,0)[lb]{\smash{=}}}%
    \put(0.50476101,0.43993779){\makebox(0,0)[lb]{\smash{+}}}%
    \put(0.01570964,0.08402322){\makebox(0,0)[lb]{\smash{+}}}%
    \put(0.50920483,0.29315906){\makebox(0,0)[lb]{\smash{0}}}%
    \put(0.45793641,0.3864741){\color[rgb]{0,0,0}\makebox(0,0)[b]{\smash{$\f{\frac{1}{\sqrt{2}}}$}}}%
    \put(0.15220108,0.45046522){\color[rgb]{0,0,0}\makebox(0,0)[b]{\smash{$\f{\sf{H}}$}}}%
    \put(0.37261491,0.36514372){\makebox(0,0)[lb]{\smash{=}}}%
    \put(0.73878628,0.36514372){\makebox(0,0)[lb]{\smash{=}}}%
    \put(0.30151367,0.08429385){\makebox(0,0)[lb]{\smash{=}}}%
    \put(0.82055267,0.3864741){\color[rgb]{0,0,0}\makebox(0,0)[b]{\smash{$\f{\frac{1}{\sqrt{2}}}$}}}%
  \end{picture}%
\endgroup%

\ee
Similarly, one can use the circuit~\eqref{eq:bell-circuit} to
generate the epsilon state.
Let us denote the Pauli matrices by $X:=\ket{0}\bra{1}+\ket{1}\bra{0}$, $Y:=-i\ket{0}\bra{1}+i\ket{1}\bra{0}$ and $Z:=\ket{0}\bra{0}-\ket{1}\bra{1}$. 
The $Z$~gate commutes with the \COPY{} tensor, and the $X$ or NOT gate (which we denote with~$\oplus$) commutes with \XOR{}.
Commuting those tensors to the right hand side,
allows us to apply Eq.~\eqref{cnot-bell-states}.  Making use of the Pauli algebra identity $ZX=iY$, one recovers the epsilon state:
\be
\renewcommand{\fscale}{1} \def\svgwidth{0.6\textwidth} 
\begingroup%
  \makeatletter%
  \providecommand\color[2][]{%
    \errmessage{(Inkscape) Color is used for the text in Inkscape, but the package 'color.sty' is not loaded}%
    \renewcommand\color[2][]{}%
  }%
  \providecommand\transparent[1]{%
    \errmessage{(Inkscape) Transparency is used (non-zero) for the text in Inkscape, but the package 'transparent.sty' is not loaded}%
    \renewcommand\transparent[1]{}%
  }%
  \providecommand\rotatebox[2]{#2}%
  \ifx\svgwidth\undefined%
    \setlength{\unitlength}{264.28125bp}%
    \ifx\svgscale\undefined%
      \relax%
    \else%
      \setlength{\unitlength}{\unitlength * \real{\svgscale}}%
    \fi%
  \else%
    \setlength{\unitlength}{\svgwidth}%
  \fi%
  \global\let\svgwidth\undefined%
  \global\let\svgscale\undefined%
  \makeatother%
  \begin{picture}(1,0.41338536)%
    \put(0,0){\includegraphics[width=\unitlength]{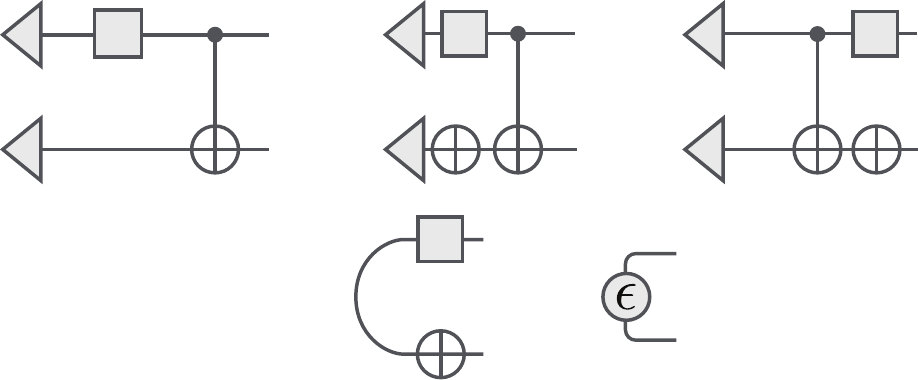}}%
    \put(0.03060187,0.37552324){\makebox(0,0)[b]{\smash{\f{1}}}}%
    \put(0.12902387,0.37675298){\makebox(0,0)[b]{\smash{\f{H}}}}%
    \put(0.3079579,0.30172573){\makebox(0,0)[lb]{\smash{=}}}%
    \put(0.44531158,0.37675298){\makebox(0,0)[b]{\smash{\f{+}}}}%
    \put(0.77231989,0.37668087){\makebox(0,0)[b]{\smash{\f{+}}}}%
    \put(0.03060187,0.25094004){\makebox(0,0)[b]{\smash{\f{1}}}}%
    \put(0.44714444,0.25094004){\makebox(0,0)[b]{\smash{\f{0}}}}%
    \put(0.77303063,0.25094004){\makebox(0,0)[b]{\smash{\f{0}}}}%
    \put(0.38477001,0.31621142){\color[rgb]{0,0,0}\makebox(0,0)[b]{\smash{\f{\frac{1}{\sqrt{2}}}}}}%
    \put(0.64207166,0.30172573){\makebox(0,0)[lb]{\smash{=}}}%
    \put(0.71888378,0.31621142){\color[rgb]{0,0,0}\makebox(0,0)[b]{\smash{\f{\frac{1}{\sqrt{2}}}}}}%
    \put(0.26974104,0.08009933){\makebox(0,0)[lb]{\smash{=}}}%
    \put(0.34655316,0.09458501){\color[rgb]{0,0,0}\makebox(0,0)[b]{\smash{\f{\frac{1}{\sqrt{2}}}}}}%
    \put(0.54823223,0.08009933){\makebox(0,0)[lb]{\smash{=}}}%
    \put(0.62504434,0.09458501){\color[rgb]{0,0,0}\makebox(0,0)[b]{\smash{\f{\frac{1}{\sqrt{2}}}}}}%
    \put(0.50585314,0.37675298){\makebox(0,0)[b]{\smash{\f{Z}}}}%
    \put(0.95386068,0.37675298){\makebox(0,0)[b]{\smash{\f{Z}}}}%
    \put(0.4799643,0.15274921){\makebox(0,0)[b]{\smash{\f{Z}}}}%
  \end{picture}%
\endgroup%

\ee
\end{example}

\begin{example}[Map-state duality]
The index raising and lowering using cups and caps can be interpreted
as a linear map between bipartite vectors and linear maps leading to a relationship between quantum states and operators acting on them.
We will start with the linear map~$A$. Raising the second index using a cup~(a)
yields a tensor with two output legs~(b),
i.e.~something that can be interpreted as a bipartite vector~$\ket{A}$.
\begin{equation}\label{fig:map-state}
\includegraphics{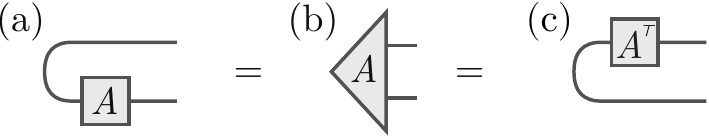}
\end{equation}
Finally, using the snake equation~\eqref{eqn:snake}
together with the equality~\eqref{eq:transpose}
we can see that this is equal to (c) the transposed map~$A^\top$
with a cup raising the input index to the other side.
Indeed, a tensor may be moved around a cup or a cap by transposing it.
The relationship \eqref{fig:map-state}
arises in practice in the following scenario from quantum information science:
\begin{equation}\label{fig:choi}
 \includegraphics{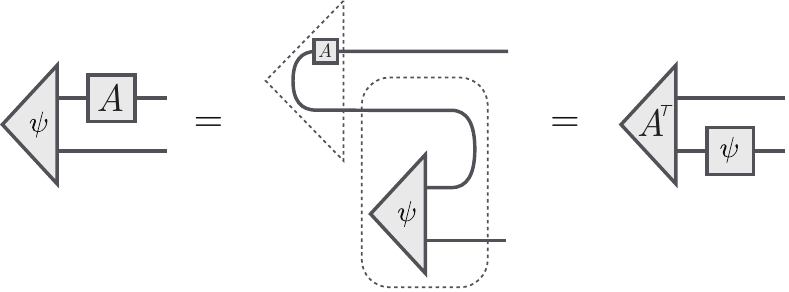} 
\end{equation}
Here an entangled state~$\ket{\psi}$ acted on by a map~$A$ can instead be viewed as a map~$\psi$ acting on a state~$\ket{A^\top}$.\footnote{
Note that the transpose in $\ket{A^\top}$ is purely to adhere to the convention in Eq.~\eqref{fig:map-state}.
In terms of numerics, the vector~$\ket{A}$ is defined here as the matrix~$A$ vectorized columnwise.
Consequently $\ket{A^\top}$ is the same matrix vectorized row-wise.}
This is a diagrammatic form of map-state duality underlying bipartite entanglement evolution \cite{Wood:2015:TNG:2871422.2871425,MB12}.  See e.g.~the survey \cite{MAL-059} which includes a detailed discussion on reshaping tensors.  
\end{example}


\paragraph{{\bf Dagger and complex conjugation.}}
\label{par:dagger}

If the order-$(p,q)$ tensor~$T$ is a map between Hilbert spaces, we may define its (Hermitian) adjoint~$T^\dagger$, an order-$(q,p)$ tensor, using the inner product:
$\inprod{x}{Ty} = \inprod{T^\dagger x}{y}$ for all~$x,y$.
Diagrammatically the adjoint is obtained by mirroring the tensor network such that input and output wires switch places,
the relative order of the tensors in the diagram is reversed, and each tensor symbol is decorated with a dagger (with $T^{\dagger \dagger} = T$).

Similarly, the dagger operation maps Hilbert space ket vectors one-to-one to their dual bra vectors (and vice versa)
in the sense of the Riesz representation theorem.
This is denoted~$\ket{a}^\dagger = \bra{a}$.
Note that we have been using this notation implicitly, 
as it should be familiar to many readers from basic quantum mechanics.

The dagger is an antilinear (or conjugate-linear) operation, $(cT+dU)^\dagger = \overline{c} \, T^\dagger+\overline{d} \, U^\dagger$,
since we have to take the complex conjugate of the scalars~$c$ and $d$.
This means that it cannot be represented by the linear cups and caps alone, unlike the transpose.
However, it can be represented as transpose together with complex conjugation in the same basis.
This is summarized in the following \emph{adjoint square}.
\be
 \includegraphics{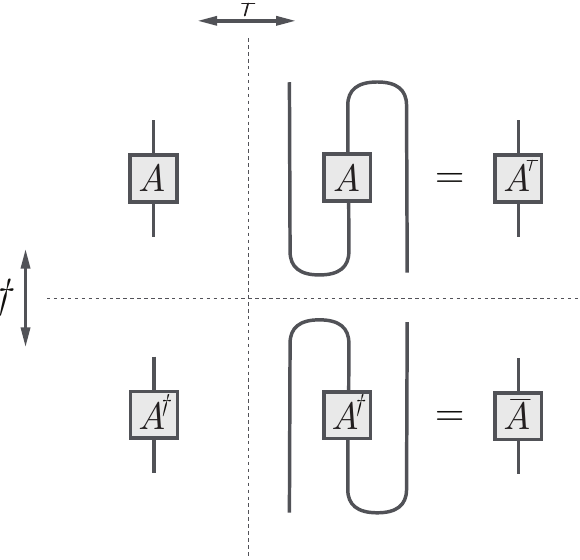}
\ee
Note that some authors mark one of the corners of each tensor box, and use the convention that mirroring a diagram across the horizontal plane corresponds to the $\dagger$ operation.
We do not adhere to this convention and instead place the dagger on the symbol such as $A\rightarrow A^\dagger$.  

For kets and bras, bending a wire on a ket yields the complex conjugate of the corresponding bra, and vice versa.
\be
 \includegraphics{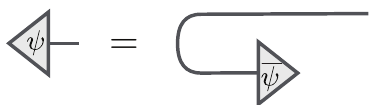}
\ee

\paragraph{{\bf Index position.}}
Now we will consider the set of operations formed by bending tensor
wires forwards and backwards using cups and caps,
as well as exchanging the order of wires using \swap{}.
If one conceptualizes a tensor as an array of numbers,
these transforms correspond to array reshapes and reorderings.
As the snake equation~\eqref{eqn:snake} shows, the action of cups and caps can be
inverted, and \swap{} is self inverse \eqref{fig:sym}. This means that all possible configurations of a tensor's wires obtained using these operations are isomorphic.

As an example, given a tensor $T\indices{^{i}_{j}}$ one can use cups and caps to naively rearrange the index elevations and positions, arriving at
\be 
T\indices{^{i}_{j}}, ~ T^{ij}, ~ T_{ij}, ~ T\indices{_{i}^{j}},
~ T\indices{^{j}_{i}}, ~  T^{ji}, ~ T_{ji} ~ \text{and} ~ T\indices{_{j}^{i}},
\ee 
for a total of eight possible reshapes. 
If the tensor has more than two indices, one can additionally use \swap{}s to
arrange the indices in any relative order.
Thus one might think that for a general $n$-index tensor
there are $n! \cdot 2^n$ different ways of arranging the indices
($n!$ different permutations of the indices, with each index being either up or down).
However, this way one overcounts the number of index configurations that are truly different.
In our example, in fact
$T\indices{^{i}_{j}} = T\indices{_{j}^{i}}$
and
$T\indices{_{i}^{j}} = T\indices{^{j}_{i}}$,
as can be seen from the diagram below:
\be
 \includegraphics{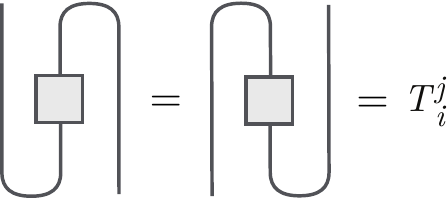}
\ee
Consequently, the tensor $T\indices{^{i}_{j}}$
in actuality only has six unique reshapes:
\be
 \includegraphics{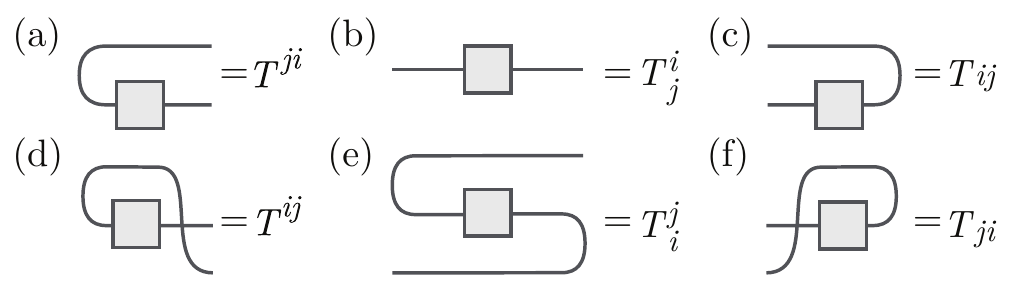}
\ee
More generally, one finds that the number of unique
reshapes generated by cups, caps and \swap{}s for an \mbox{order-$(p,q)$} tensor
$T\indices{^{i_1\cdots i_p}_{j_1 \cdots j_q}}$
is $(p+q+1)!$. 

\section{Diagrammatic SVD}\label{sec:diagrammaticSVD}

In this section, we explain the diagrammatic version of
the singular value decomposition~(SVD).  This method is at the heart of many
numerical simulation algorithms in wide use today---we are explaining it as a prelude leading to Section~\ref{sec:dmps}.
The~SVD factors an arbitrary order-$(1,1)$ tensor into well defined building blocks
with simple properties:
(i) an order-$(1,1)$ diagonal tensor storing the singular values, and
(ii) two order-$(1,1)$ unitary tensors.
As several tensor legs can always be grouped together to form a single leg, the method works for any tensor of order two or higher.

Interpreting order-$(1,1)$ tensors as linear maps (or simply as matrices),
we may use the SVD to factor
any tensor $T: A \to B$ (for vector spaces $A$ and $B$) as
\be
T\indices{^b_a} = U\indices{^b_j} \Sigma\indices{^j_i} V\indices{^i_a},
\ee
where $U$~and~$V$ are unitary,
and $\Sigma$ is 
real, non-negative, and diagonal in the computational basis.
$\Sigma$ has the singular values $\{\sigma_k\}_k$ of~$T$ on its diagonal,
typically arranged in a nonincreasing order:
$\sigma_1 \ge \sigma_2 \ge \ldots \ge \sigma_{\min(\dim A, \dim B)} \ge 0$.
It can be expanded as
\be 
\Sigma
= \sum_{k} \sigma_k \ket{k}_B \bra{k}_A.
\ee
Diagrammatically this is represented as
\be
\includegraphics{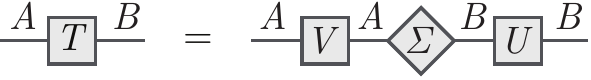}\label{fig:SVD-s}.
\ee

As will be seen in the next section, the SVD is centrally employed in the efficient representation of certain quantum states by tensor networks.  The idea is to represent a small but physically relevant portion of the Hilbert space---such as low entanglement states---by repeated application of the SVD paired with low-rank approximations.

The rank of a matrix~$T$ is the number of non-zero singular values it has.
To determine its optimal rank-$r$ approximation (with $r < \rank(T)$), we can turn to a classic theorem by Eckart and Young which was generalized by Mirsky.
Given the SVD $T = U\Sigma V^\dagger$, we will discard $\rank(T)-r$ smallest singular values in~$\Sigma$ by setting them to zero, obtaining~$\Sigma'$.
This process is often called trimming.

This gives rise to $T' = U\Sigma' V^\dagger$, an approximation of~$T$.  The Eckart-Young-Mirsky theorem states that
\begin{equation}
\lVert T -T' \rVert =  \min_{\rank(\hat{T}) \le r} \lVert T -\hat{T}\rVert
\end{equation}
for any unitarily invariant matrix norm.\footnote{Interested readers can try to get their hands on copies of
C. Eckart and G. Young, ``The approximation of one matrix by another of lower rank,'' Psychometrika 1, (1936)
and
L. Mirsky, ``Symmetric gauge functions and unitarily invariant norms,'' The Quarterly Journal of Mathematics 11:1, 50--59 (1960).}
Here $\hat{T}$ is any approximation to $T$ of the same or lesser rank as~$T'$.
This implies that truncating or trimming $\Sigma$ in this way yields as good of an approximation as one can expect.
In the following section, we will specifically consider the induced error for such an approximation.


Using the wire bending techniques from Section~\ref{sec:wires},
we immediately obtain the Schmidt decomposition as a corollary to the SVD:
\be
\label{fig:SVD-Schmidt2}
\includegraphics{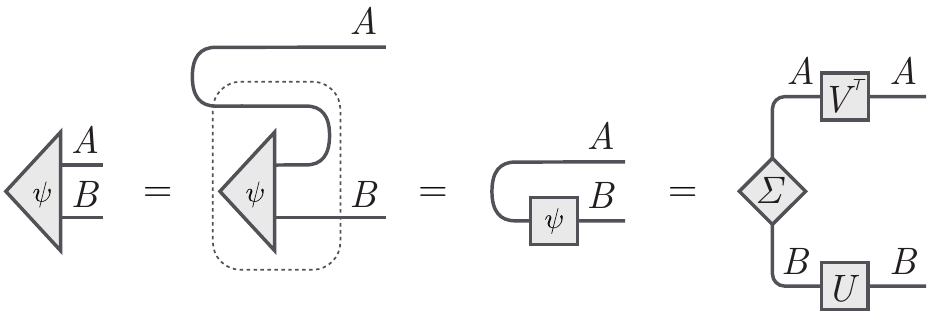}
\ee
Given a vector $\ket{\psi} \in A \otimes B$ (for example a ket vector
describing a pure state of a bipartite quantum system),
we may use the snake equation to
convert it into a linear map~$\psi:A \to B$ (inside the dashed region). 
Now we apply the SVD on $\psi$ as above.
Diagram reorganization leads to the diagrammatic Schmidt decomposition
\be
\ket{\psi}_{A\otimes B} = \sum_i \sigma_i \ket{\varphi_i}_A\ket{\phi_i}_B.
\ee
The singular values $\{\sigma_k\}_k$ now correspond to the Schmidt coefficients.
If $\ket{\psi}$ is normalized, we have $\sum_k \sigma_k^2 = 1$.

\begin{example}[Entanglement topology]
\label{example:entanglement-topology}
The topology of the bipartite quantum state
$\ket{\psi}$
depends solely on the Schmidt coefficients~$\{\sigma_k\}_k$.
\be
\includegraphics{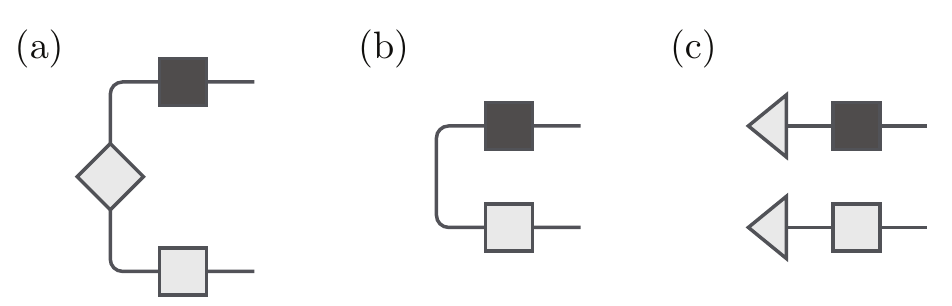}
\ee
If the coefficients are all equal, we may
replace the diamond in (a) with a cup tensor times a scaling factor
as in diagram~(b).
This corresponds to a maximally entangled state.
In the other extreme, illustrated in diagram (c), we have just one
nonzero Schmidt coefficient $\sigma_1 = 1$. In this case the diagram
breaks into two pieces and thus corresponds to a factorizable state. The number of nonzero Schmidt coefficients is called the \emph{Schmidt rank}
of the decomposition---see Def.~\ref{def:entropy} for the relation with R{\'e}nyi entropy of order zero.
\end{example}


\begin{example}[Concurrence---part II] \label{ex:con-2}
Continuing on from Example \ref{ex:concurrence}, one can apply the Schmidt decomposition to the
tensor network defining the concurrence of a two-qubit state~$\ket{\psi}$. We obtain
\be
\includegraphics[width=0.9\textwidth]{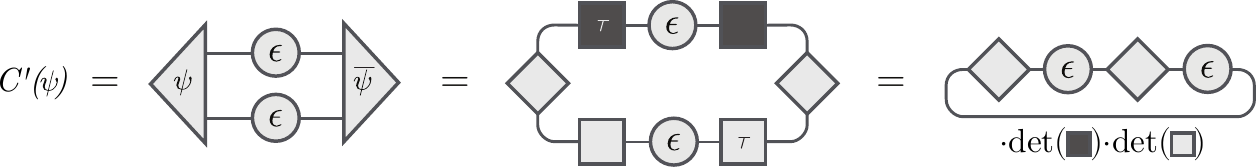}
\ee
and thus
\be
C(\psi) = |C'(\psi)| =
\left|\trace((\epsilon \Sigma)^2)\right|
= 2 \sigma_1 \sigma_2
= 2 \, |\det(\psi)|.
\ee
The unitaries $\blacksquare$ and~$\square$ vanish from the expression since $|\det(\blacksquare)| = |\det(\square)| = 1$.
Given the basis expansion of the state,
\begin{equation}
\ket{\psi} = a\ket{00} +b\ket{01} +c\ket{10} +d\ket{11},
\end{equation}
its two Schmidt coefficients are given as
\begin{equation}
\sigma_k^2=\frac{1}{2}\left( 1 +(-1)^{k+1} \sqrt{1-4|ad-bc|^2}
\right).
\end{equation}
\end{example}

\begin{example}[Purification backwards] 
 
For any bipartite ket vector~$\ket{\psi}$,
the partial trace of $\ketbra{\psi}{\psi}$ yields a positive semidefinite operator.
We can see this by using the Schmidt decomposition, and then reorganizing the diagram
so that two of the unitaries cancel:
\be
\includegraphics{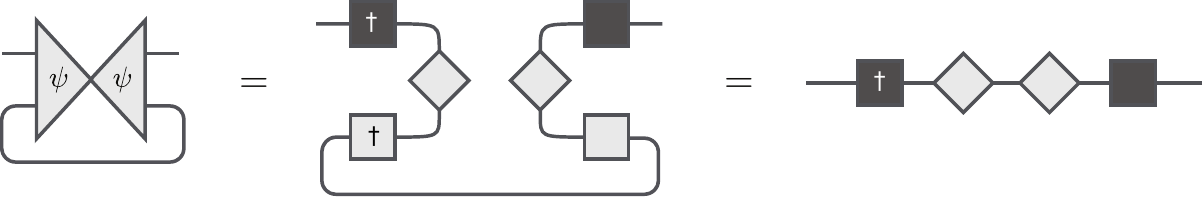}
\ee
On the right hand side we have an eigendecomposition of a square matrix with
strictly nonnegative eigenvalues~$\{\sigma_k^2\}_k$,
which can be interpreted as a density matrix if $\ket{\psi}$ is normalized.
Conversely, any density matrix representing a mixed quantum state
can be \emph{purified}, or expressed as the partial trace of a bipartite pure state.
\end{example}

\section{Matrix Product States}\label{sec:dmps}

Matrix product states (MPSs) are
quantum states presented as a linear chain or ring of tensors.
Any quantum state can be exactly represented in this form and
the representation is known to approximate a class of 1D gapped systems efficiently \cite{2007JSMTE..08...24H}.   We will explain the basic ideas of the MPS representation here and point the reader to~\cite{MPSreview08,TNSreview09} and the references therein for additional information.

Given an $n$-party quantum state $\ket{\psi}$, fully describing this state generally requires an
amount of information (or computer memory) that grows exponentially with~$n$.
If $\ket{\psi}$ represents the state of $n$~qubits,
\be 
\ket{\psi}= \sum_{i j \cdots k}\psi_{i j \cdots k}\ket{i j \cdots k},
\ee 
the number of independent coefficients $\psi_{i j \cdots k}$ in the basis expansion in general would be $2^n$
which quickly grows into a computationally unmanageable number as~$n$ increases.
The goal is to find an alternative representation of~$\ket{\psi}$ which is less data-intensive.
We wish to write $\ket{\psi}$ as 
\begin{equation}\label{eqn:matrix}
\ket{\psi} = \sum_{i j \cdots k} \trace(A_{i}^{[1]} A_{j}^{[2]} \cdots A_{k}^{[n]}) \ket{i j \cdots k},
\end{equation}
where $A_{i}^{[1]}, A_{j}^{[2]}, \ldots, A_{k}^{[n]}$ are indexed sets of matrices.
Calculating the components of $\ket{\psi}$ then becomes a matter of calculating the products of matrices,
hence the name \emph{matrix product state}.

If the matrices are bounded in size, the representation becomes efficient
in the sense that the amount of information required to describe them is only linear in~$n$.
The point of the method is to choose these matrices such that they provide a good (and compact) approximation to~$\ket{\psi}$.
For instance, if the matrices are at most $\chi$ by $\chi$, the size of the representation scales as~$n d \chi^2$,
where $d$~is the dimension of each subsystem.


Without loss of generality, we will now show how to obtain an MPS representation of an arbitrary four-party state~$\ket{\psi}$.
The key ingredient is the recursive application of the singular value decomposition (SVD) presented in Section~\ref{sec:diagrammaticSVD}.
We start by considering the tensor which represents the state vector.
We select a bipartition that separates one leg from the rest (starting at either end), and apply the SVD.
This process is then repeated, traversing the entire tensor.  This results in a 1D tensor network representation of the
state, as shown below.
\be
\includegraphics{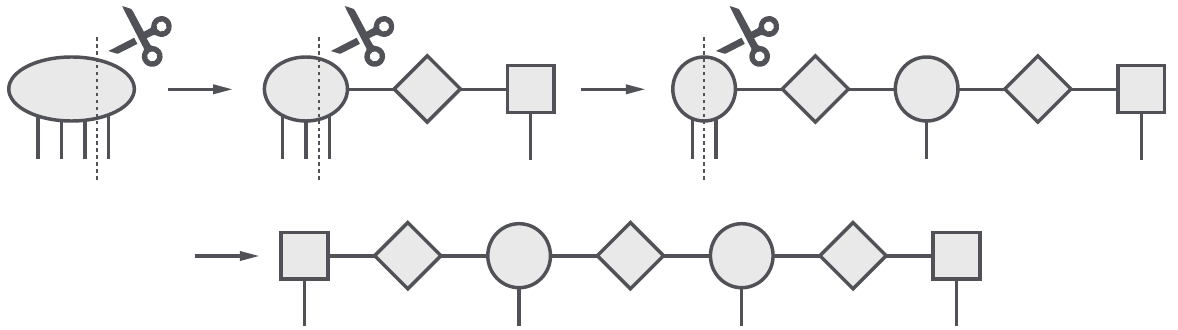}
\ee
Finally the tensors are grouped---though the grouping has some ambiguity---resulting in the typical form shown below.  
\be
\label{eq:mps}
\renewcommand{\fscale}{0.8} \def\svgwidth{0.8\textwidth} 
\begingroup%
  \makeatletter%
  \providecommand\color[2][]{%
    \errmessage{(Inkscape) Color is used for the text in Inkscape, but the package 'color.sty' is not loaded}%
    \renewcommand\color[2][]{}%
  }%
  \providecommand\transparent[1]{%
    \errmessage{(Inkscape) Transparency is used (non-zero) for the text in Inkscape, but the package 'transparent.sty' is not loaded}%
    \renewcommand\transparent[1]{}%
  }%
  \providecommand\rotatebox[2]{#2}%
  \ifx\svgwidth\undefined%
    \setlength{\unitlength}{366.48360001bp}%
    \ifx\svgscale\undefined%
      \relax%
    \else%
      \setlength{\unitlength}{\unitlength * \real{\svgscale}}%
    \fi%
  \else%
    \setlength{\unitlength}{\svgwidth}%
  \fi%
  \global\let\svgwidth\undefined%
  \global\let\svgscale\undefined%
  \makeatother%
  \begin{picture}(1,0.09226197)%
    \put(0,0){\includegraphics[width=\unitlength]{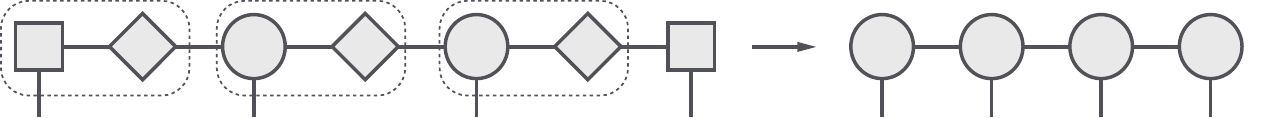}}%
    \put(0.69472757,0.05530603){\color[rgb]{0,0,0}\makebox(0,0)[b]{\smash{$\f{A^{[1]}}$}}}%
    \put(0.77986094,0.05530603){\color[rgb]{0,0,0}\makebox(0,0)[b]{\smash{$\f{A^{[2]}}$}}}%
    \put(0.86717726,0.05530603){\color[rgb]{0,0,0}\makebox(0,0)[b]{\smash{$\f{A^{[3]}}$}}}%
    \put(0.95231071,0.05530603){\color[rgb]{0,0,0}\makebox(0,0)[b]{\smash{$\f{A^{[4]}}$}}}%
  \end{picture}%
\endgroup%

\ee
We note that one can recover the form in Eq.~\eqref{eqn:matrix} as
\begin{equation}\label{eqn:MPS}
\ket{\psi} = \sum_{ijkm} A^{[1]}_i A^{[2]}_j A^{[3]}_k A^{[4]}_m\ket{ijkm}.
\end{equation}
In this case the first and last sets of matrices $A^{[1]}_i$ and $A^{[4]}_m$ have just a single row and a single column, respectively,
so the product always yields a scalar and the trace is not required. 
The tensor network in Eq.~\eqref{eq:mps}
is called an MPS with open boundary conditions.
MPSs also come with periodic boundary conditions, in which case the tensors form a ring instead of a chain,
and the trace represents the contraction that closes the ring.

You might have noticed that we made a choice to perform the
factorization starting from the left of the tensor and applying the SVD
successively on tensors as we moved to the right.  This apparent
ambiguity has been characterized in detail~\cite{2006quant.ph..8197P}. 
 For open boundary conditions as have been considered here, there is a canonical choice unique up to degeneracies in
the spectrum of local reduced density operators~\cite{2006quant.ph..8197P}.


For a general $n$-qubit quantum state it can be shown that the MPS matrix size
can be bounded by $\chi = 2^{\lfloor n/2\rfloor}$, which still grows exponentially
as expected.
A compact approximate representation is obtained by choosing a cutoff
value~$\xi$ for the singular values across each partition, or a maximum
number~$\chi$ of singular values to be kept---see Example \ref{ex:approx}.  This allows one to compress data
by truncating the Hilbert space and is at the heart of the MPS algorithms.
If an MPS is cut into two parts, the Schmidt rank of the decomposition, describing the degree of entanglement between the parts,
is always $\le \chi$, the dimension of the internal wire that was cut---see Example \ref{example:entanglement-topology} for Schmidt rank and set $q=0$ in Eq.~\eqref{eqn:renyi} for the connection to entropy.

Above we have illustrated how to obtain the MPS representation of any pure quantum state,
but it is normally not practical to factor states in this way for computational reasons.
Instead, efficient MPS-generating algorithms are given an indirect, compact description of a state e.g.~in the form of a nearest-neighbor Hamiltonian whose ground state we are interested in, and
they then iteratively produce an MPS that closely approximates that state.
The seminal algorithm of this type is the Density Matrix Renormalization Group (DMRG),
which essentially works as a variational method in MPS space.
Another class of algorithms can efficiently time-evolve MPSs under
a nearest-neighbor Hamiltonian.
One of the most used methods of this type is Time-Evolving Block Decimation (TEBD).

\begin{example}[MPS approximation error]\label{ex:approx}

As explained, matrix product state algorithms employ repeated application of the singular value decomposition.  The size of the representation can be reduced by lossy truncation in one of two ways.  Each of these rely on truncation of singular values.  For a fixed rank, the Eckart-Young-Mirsky theorem---from the last Section---tells us that truncation of the singular values is the best approximation that one can expect. 

In the first approximation, one can
simply discard some fixed number of lowest singular values and their corresponding vectors---in other words, we will fix the dimension $\chi$ of all internal wires.
In another approximation---which we will consider here---one will pick a cutoff value~$\xi$,
and truncate all singular values which are less than this.
Such a cutoff value is not guaranteed to provide a useful partition of the singular values---e.g.~it could be smaller than the smallest singular value.  
Here we will analyze the errors of this truncation assuming this cutoff partitions the singular values---which in practice is very often the case.  

Given a bipartite state $\ket{\psi}$ we write
\begin{equation}\label{eqn:svd1}
  \ket{\psi} = \sum_{i=1}^k \sigma_i \ket{u_i}\ket{v_i} = \sum_{i=1}^k \sigma_i \ket{i}.
\end{equation}
where to simplify the notation we write $\ket{u_i}\ket{v_i}$ as just $\ket{i}$.
We order the singular values in a non-decreasing sequence and introduce a cutoff $\xi > 0$:
\begin{equation}
0 < \sigma_1\leq \sigma_2\leq \sigma_3\leq\cdots \le \sigma_n\leq \xi \leq \sigma_{n+1}\leq\cdots \leq \sigma_k.
\end{equation}
As a heuristic, the small cutoff $\xi$ can be chosen such that $\xi < \frac{1}{\sqrt{\dim H}}$ where $H$ is the vector space acted on by $\ket{\psi}$---you could also write $\xi < \frac{1}{\sqrt{d^{q}}}$ where $d$ is the dimension of $q$ constituent spaces.   

And then we will partition our space in terms of the $n$ singular values less than $\xi$ and the $k-n$ ones that are greater than $\xi$. 
\begin{equation}\label{eqn:psi}
 \ket{\psi}= \sum_{i=1}^n \sigma_i \ket{i} + \sum_{j=n+1}^k \sigma_j \ket{j} = \sum_{i=1}^n \sigma_i \ket{i} + \ket{\psi'},
\end{equation}
where $\ket{\psi'}$ in \eqref{eqn:psi} represents a (so far, non-normalized) approximation to $\ket{\psi}$. To understand the limits of validity of this approximation, we consider the inequality  
\begin{equation}
\label{eq:sigma_ineq}
 \sum_{i=1}^n \sigma_i \leq  n\cdot  \sigma_n \leq n \cdot \xi,
\end{equation}
and hence $n \cdot \sigma_n^2 \leq n \cdot \xi^2$.  

Provided $\xi$ is small, and for constant $n$ we can return to \eqref{eqn:psi} and consider another normalized but otherwise arbitrary vector $\ket{\phi}$
(element of the same space as $\ket{\psi}$ and $\ket{\psi'}$)
\begin{equation}\label{eqn:psiap}
|\braket{\phi}{\psi - \psi'}|
= \left| \sum_{i=1}^n \sigma_i \braket{\phi}{i} \right|
\le \sum_{i=1}^n \sigma_i \lvert\braket{\phi}{i}\arrowvert
\leq \sum_{i=1}^n \sigma_i \leq n\cdot \xi,
\end{equation}
which scales linearly in the upper-bound of the error independent of $\ket{\phi}$.  
%

It's common to introduce a function $\mathcal{O}(\xi)$ to collect terms up to a constant that approach zero no slower than $\mathcal{O}$'s argument.  So \eqref{eqn:psiap} simply says that the absolute value of the difference between $\ket{\psi'}$ and $\ket{\psi}$ projected onto $\bra{\phi}$ is bounded by a function proportional to $\xi$.  From this we can readily conclude that the error in the approximation considered here is linear in $\xi$---or in other words, the error scales at most as $\mathcal{O}(\xi)$. 

However, as $\ket{\psi}$ is normalized, we note that 
\begin{equation}
 \braket{\psi}{\psi} = \sum_{i=1}^n \sigma_i^2 + \sum_{j=n+1}^k \sigma_j^2 = 1,
\end{equation}
and then calculate the inner product  
\begin{equation}
  \braket{\psi}{\psi'} = \sum_{j=n+1}^k \sigma_j^2 = 1 - \sum_{i=1}^n \sigma_i^2 \geq 1 - n \xi^2 = 1+ \mathcal{O}(\xi^2).
\end{equation}
We hence conclude that the error in the inner product is only quadratic.
In fact, if we normalize $\ket{\psi'}\rightarrow\ket{\psi''}$ and calculate 
\begin{equation}
 |\braket{\psi}{\psi''}|^2 = 1 - \sum_{i=1}^n \sigma_i^2 \geq 1 - n \xi^2 = 1+ \mathcal{O}(\xi^2),
\end{equation}
we recover the same thing. 
\end{example}

\begin{example}[MPS for the GHZ state]\label{ex:mps-ghz}
The standard MPS representation of the Greenberger-Horne-Zeilinger (GHZ) state is given as 
\be\label{eqn:ghzMPS}
\ket{\text{GHZ}} = \frac{1}{\sqrt{2}}\trace\left( \begin{array}{cc}
\ket{0} & 0 \\
0 & \ket{1} \end{array} \right)^{n}=
\frac{1}{\sqrt{2}}(\ket{00\ldots 0}+\ket{11\ldots 1}).
\ee
Alternatively, we may use a quantum circuit made of \cnot{} gates to construct
the GHZ state,
and then use the rewrite rules employed in Examples \ref{ex:circuits-2} and \ref{ex:state-prep}
to recover the familiar MPS comb-like structure consisting of \COPY{} tensors:
\be
\includegraphics{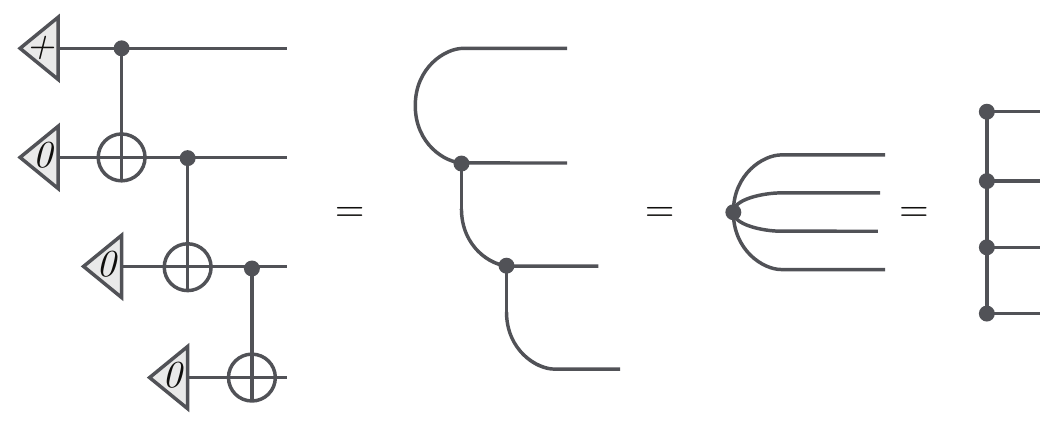}
\ee 
Diagrammatically, any tensor network formed from connected \COPY{}-tensors reduces to a single dot
with the appropriate number of input and output legs.
Hence one might write the n-party GHZ-state as 
\be 
\ket{\text{GHZ}} = \frac{1}{\sqrt{2}}\sum_{ijk\ldots l} \COPY^{ijk\ldots l}\ket{ijk\ldots l}.
\ee 
\end{example}


\begin{example}[MPS for the W state]\label{ex:w-state}
Like the GHZ state from Example \ref{ex:mps-ghz}, the $n$-qubit W~state ($n\geq 3$) has the following MPS representation:
\be\label{eqn:wnMPS}
\ket{W}
=
\frac{1}{\sqrt{n}}
\begin{pmatrix} \ket{1} & \ket{0} \end{pmatrix}
\left( \begin{array}{cc}
\ket{0} & 0 \\
\ket{1} & \ket{0} \end{array} \right)^{n-2}
\begin{pmatrix} \ket{0}\\ \ket{1} \end{pmatrix}
=
\frac{1}{\sqrt{n}} \left(
\ket{10\ldots 0}+\ket{010\ldots 0}+\ldots +\ket{0\ldots 01}
\right).
\ee
\end{example}

\begin{example}[The AKLT model] \label{ex:AKLT} 
The AKLT model~\cite{1987PhRvL..59..799A} (named after the authors Affleck, Kennedy, Lieb and Tasaki)
is a theoretically important exactly solvable model of a spin-1 Heisenberg chain with an extra
quadratic interaction term:
\be
H = \sum_j \vec{S}_j \cdot \vec{S}_{j+1} + \frac{1}{3} (\vec{S}_j \cdot \vec{S}_{j+1})^2,
\ee
where $\vec{S}$ is a three-vector of the familiar spin-1 operators.
The exact ground state of this Hamiltonian
has an elegant expression as a matrix product state.
Here we will carry on from Examples~\ref{ex:epsilon} and~\ref{ex:spinor}
which define and use the $\epsilon$ tensor---see also Example \ref{ex:state-prep} which provides a quantum circuit realization for the corresponding singlet state.
We start with a tensor product of these states,
\be 
\includegraphics{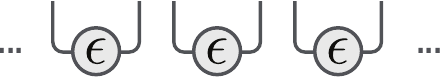}
\ee
and then project each neighboring qubit pair onto a three-dimensional (spin-1) Hilbert space using the projectors~$P$:
\be
\includegraphics{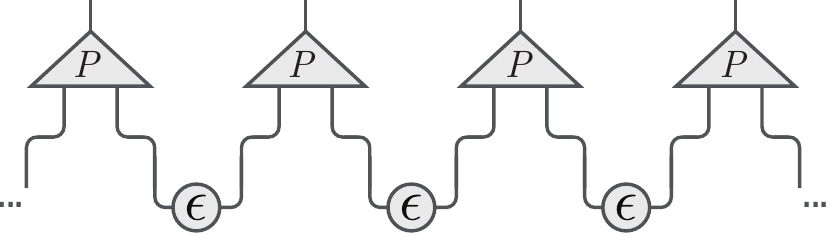}
\ee
 
The open wires on the $P$'s have three degrees of freedom (spin-1).  The projectors are defined as 
\be \label{eqn:projaklt}
P = \ket{\textbf{+1}}\bra{11} + \frac{1}{\sqrt{2}} \ket{\textbf{0}}(\bra{01}+\bra{10})+\ket{\textbf{--1}}\bra{00}.
 \ee
The kets $\ket{\textbf{+1}}$, $\ket{\textbf{0}}$ and $\ket{\textbf{--1}}$ are the standard spin-1 basis states and clearly the bond dimension $\chi=2$.
The AKLT state obtained after projection is rotationally symmetric but has a non-trivial entanglement structure and a host of other interesting properties---see e.g.~\cite{2014AnPhy.349..117O} and the references therein.

\end{example}

The singular values found from the MPS factorization can be used to form a complete polynomial basis to express invariant quantities related to an MPS.
This has a close connection to Schmidt rank and other concepts---see Examples~\ref{example:entanglement-topology} and~\ref{ex:con-2}.

\begin{definition}[R{\'e}nyi and von Neumann  entropies]\label{def:entropy}
Given a density operator~$\rho$, its R{\'e}nyi entropy of order~$q$ is defined to be
\begin{equation}\label{eqn:renyi}
 H_q (\rho) = \frac{1}{1-q} \log \Tr(\rho^q) = \frac{1}{1-q} \log \sum_i
\lambda_i^q,
\end{equation}
where $\lambda_i$ are the eigenvalues of $\rho$.  
The case $q\to0$ gives the rank of~$\rho$ (here we define $0^0 = 0$),
and the limit 
$q \to 1$ recovers the familiar von Neumann entropy
\be \label{eqn:vne}
H_{q\rightarrow 1}
(\rho) = - \Tr(\rho \log \rho).
\ee 
\end{definition}
For a pure bipartite state~$\ket{\psi}$ with subsystems $A$ and~$B$,
the entropies of the reduced density operator
$\rho_A = \Tr_B(\ketbra{\psi}{\psi})$
can be used to quantify bipartite entanglement in the state.
In this context they are called entanglement entropies.
\vb{Are all of them used, or just $q=1$?}
The eigenvalues of $\rho_A$ (and~$\rho_B$) are the squares of the Schmidt coefficients of~$\ket{\psi}$, and
$H_0(\rho_A)$ is equal to the Schmidt rank of the bipartition---see Example~\ref{example:entanglement-topology}.

Area laws are quantified in terms of the scaling of entropy across tensor network partitions~\cite{2010RvMP...82..277E}.
Whenever you bipartition a tensor network representing a pure quantum state,
the total dimension~$\chi$ of wires ``cut'' by this partition gives an upper bound to the entanglement entropy:
$H_0(\rho_A) \le \chi$ and
$H_1(\rho_A) \le \log \chi$.

\section[Counting]{Counting by Tensor Contraction} 

In the tensor network language, counting problems can be expressed by evaluating fully contracted diagrams \cite{Penrose,collected,VTN,BMT15}. 

To understand counting problems, imagine a phone book.  But this phone book has a problem.  The names are arranged randomly, without the standard alphabetical order we'd all expect.  If you're task is to determine Stephen Clark's phone number, the average number of names you'll need to examine before finding `Stephen Clark' is exactly half the number in the phone book---assuming the name is unique.  This is an example of a search problem.  If we tell you the page number and location on the page where Stephen Clark's name appears, you can easily check and see if we're correct or not.  Often times in computer science, problems are classified not in terms of how hard they are to solve---which is often unknown---but in terms of how hard it is to check if a given solution is correct or not.  

Counting problems arise from search problems in the following way.  Imagine you want to determine all the entries in this random phone book with the last name `Jaksch'.  This is then the counting version of the above search problem.  

In these examples, both problems can be solved efficiently in the number of entries in the phone book.  `Efficiently' in the language of computer science means that the computational memory and runtime required is less than a polynomial in the problem size---in this case, the number of phone book entries.  

In the language of computer science, more generally search problems are complete for the complexity class {\sf NP}.  Counting versions of {\sf NP}-complete problems are {\sf \#P}-complete \cite{valiant1979complexity2} in the language of computational complexity theory (pronounced sharp-P).

Connections between counting problems and tensor networks arose from the very early days of tensor networks.
Penrose showed \cite{Penrose} that certain graph coloring problems can be solved by tensor contraction.
We're going to outline Penrose's algorithm in Section~\ref{sec:color}.
Before doing that, we will describe in Section~\ref{sec:counter} how tensor contractions can count the number of inputs that cause a given Boolean function to output~$1$.

\subsection[Boolean Tensor Counters]{Counting Boolean Formula Solutions}\label{sec:counter} 
  
A Boolean function (a.k.a. switching function) takes an $n$-bit string of binary numbers---e.g.~$00110101011$---and maps this to a single binary digit (either $0$ or $1$). Several problems in physics can be mapped to Boolean functions \cite{2012EL.....9957004W}. 	
In what follows, we will often denote this $n$-bit string as $\x$.
We will show how to associate the Boolean function $f(\x)$ with a non-normalized quantum state, written as a tensor network. We will establish the following (Remark \ref{remark:boolean}) by a few examples.  

\begin{definition}[The class of Boolean tensors \cite{VTN,BMT15}]
Every Boolean function
$f(\x)$ gives rise to a Boolean tensor
\begin{equation}\label{eqn:bool}
 f = \sum_{\x} \ketbra{f(\x)}{\x},
\end{equation}
with binary coefficients in $\{0,1\}$. 
Moreover, a tensor network representing this state is determined from the classical logic gate network description of~$f(\x)$.
\label{remark:boolean}
\end{definition}
As an example, consider the logical AND operation which takes Boolean input variables $x_1$ and $x_2$ and outputs the Boolean product $x_1\wedge x_2$---as a tensor, we write 
\begin{equation}
 \text{AND} := \sum_{x_1, x_2} \ketbra{x_1\wedge x_2}{x_1, x_2} = \ketbra{0}{00} + \ketbra{0}{01} + \ketbra{0}{10} +\ketbra{1}{11}.
\end{equation}
We will use the standard AND gate symbol also for the corresponding tensor:
\be
 \includegraphics{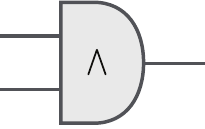}
\ee

Using \eqref{eqn:bool} we can map any Boolean tensor $f$ to an (unnormalized) Boolean state $\ket{f}$ by contracting the output with~$\ket{1}$:
\be
\label{eq:bool-state}
\ket{f} = f^\top \ket{1} =
\sum_{\x} \braket{f(\x)}{1}\ket{\x}
= \sum_{\x} f(\x) \ket{\x}.
\ee
In addition to AND we have already seen other examples from this class of states,
the GHZ and W states from Examples~\ref{ex:mps-ghz} and~\ref{ex:w-state}.
The corresponding Boolean functions (for three variables/qubits) are
\begin{equation}
f_\text{GHZ}(\x) = x_1x_2x_3 +(1-x_1)(1-x_2)(1-x_3)
\end{equation}
and
\begin{equation}
f_\text{W}(\x) =  x_1x_2(1-x_3) + x_1(1-x_2)x_3 + (1-x_1)x_2x_3.
\end{equation}
Inserting these functions in Eq.~\eqref{eq:bool-state}, we obtain the corresponding (unnormalized) quantum states:
\begin{align}
  \ket{\text{GHZ}} &= \ket{000}+\ket{111},\\
  \ket{\text{W}}   &= \ket{001}+\ket{010}+\ket{100}.
\end{align}

Now, given a Boolean function~$f(\x)$, each input for which the function returns~$1$ is said to satisfy the function.
The number of satisfying inputs is then equal to the size of the support of~$f(\x)$.
\be
  \includegraphics{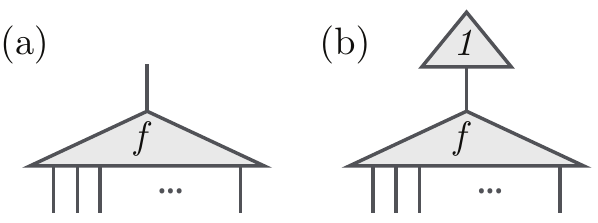}
\ee
We may count the number of inputs that satisfy~$f(\x)$
by contracting the tensor~$f$ with $\ket{1}$ to obtain the corresponding Boolean state $\ket{f}$ (diagram (b)).
The contraction post-selects the tensor network such that its support
now consists solely of inputs that satisfy~$f(\x)$.

To explicitly translate this into a counting problem, we compute the squared norm,
\be 
\|\ket{f}\|^2 = \braket{f}{f}
=\sum_{\x, \y} f(\x) f(\y) \braket{\x}{\y} = \sum_\x f(\x)^2 = \sum_\x f(\x),
\ee 
which clearly gives the number of satisfying inputs.
In general for Boolean states the square of the two-norm always equals the one-norm
since $f(\x) \in \{0,1\}$.
In diagram form this is depicted as
\be
 \includegraphics{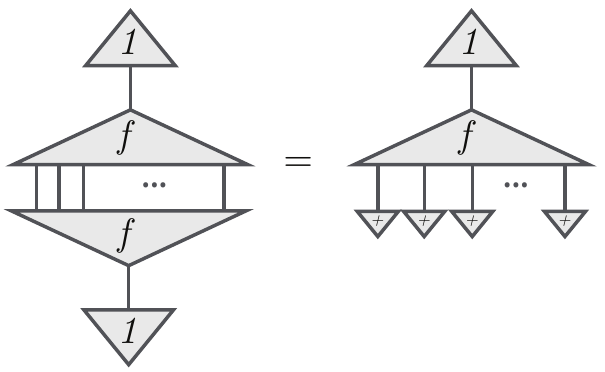}
\ee
where $\ket{+}:= \ket{0}+\ket{1}$
and thus
$\ket{+}\otimes\ket{+}\otimes \ldots \otimes \ket{+} = \sum_{\x} \ket{\x}$. More formally, 
\begin{theorem}[Counting SAT solutions \cite{VTN,BMT15}]\label{theorem:3-SAT}
Let $f$ be a SAT instance. Then the standard two-norm length squared of the corresponding Boolean state $\ket{f}$ gives the number of satisfying assignments of the problem instance. 
\end{theorem}

Solving the counting problem for general formula is known to be {\sf \#P}-complete \cite{valiant1979complexity2}. Just to remind you what we talked about in the introduction to this section, computational complexity jargon for the set of the counting problems associated with the decision problems---decision problems seek to determine if a state is satisfiable at all, whereas counting problems seek to determine the total number of satisfying solutions.  Indeed, the condition 
$
\braket{f}{f} > 0
$
implies that the SAT instance $f$ has a satisfying assignment.
Determining whether this condition holds for general Boolean states is a {\sf NP}-complete decision problem---as described in the introduction to this section.

Formulating these problems as tensor contractions allows the adaptation of tools developed to simulate quantum systems and circuits such that they now apply to an areas
traditionally considered in computer science \cite{VTN,BMT15}. We adapted these tools and discovered efficient tensor network descriptions of finite Abelian lattice gauge theories \cite{2012JPhA...45a5309D}.  These tools also lead to the discovery of a wide class of efficiently contractable tensor networks, representing counting problems \cite{BMT15}.

\begin{example}[AND from Toffoli]
The Toffoli gate has long been studied in reversible computing, and also in quantum computing.
One can view this operation as being formed internally by an AND tensor, two copy tensors (black dots)
 and one \XOR{} tensor ($\oplus$)---see Examples \ref{ex:circuits-1} and \ref{ex:mps-ghz}.  One can create the $\ket{\text{AND}}$ state in an experiment by preparing the state
\begin{equation}
 \ket{+}\ket{+}\ket{0},
\end{equation}
where $\ket{+}:= \ket{0}+\ket{1}$,
and then applying the Toffoli gate.  This is illustrated with the following tensor diagram.  
\be
 \includegraphics{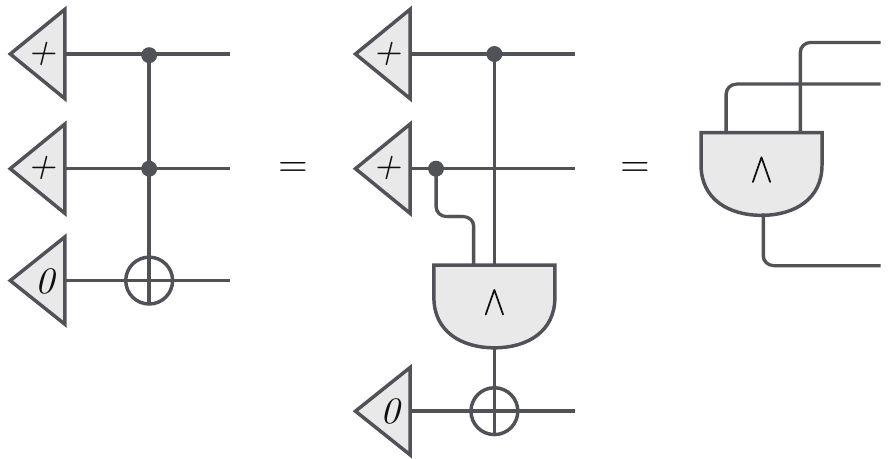}  
\ee
\end{example}

\begin{example}[Hadamard from AND]
By considering the contraction formed with the state
$\ket{-}:= \frac{1}{\sqrt{2}}(\ket{0}-\ket{1})$ and the output of the AND tensor, one recovers the Hadamard gate defined in Eq.~\eqref{eqn:hadamard}.
Note that we've been using $\ket{+}$ without normalization and here we normalize $\ket{-}$. 
\be
 \includegraphics{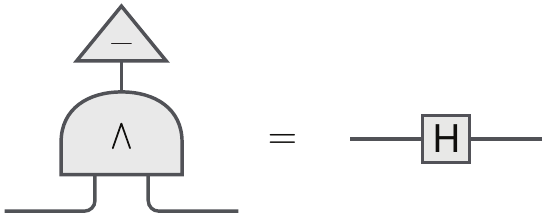}
\ee

\end{example}

\begin{example}[De Morgan's laws]
A common identity used in Boolean algebra is 
\be
\neg (a \wedge b) = (\neg a)\vee(\neg b),
\ee 
where negation denoted as $\neg$, logical OR as $\vee$, and logical AND as $\wedge$.
When expressed as a tensor network,
\be
 \includegraphics{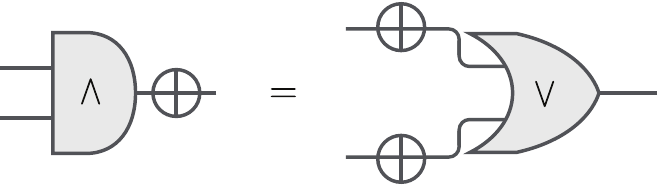}
\ee
this relation has the same structure as the relationship between the \COPY{} and \XOR{} tensors and the Hadamard gate in Eq.~\eqref{eq:copy-vs-xor}
Such diagrammatic rewrite rules can be used to formalize a system of graphical reasoning, and reduce calculations through easily employed graphical rewrite identities.
\end{example}

The synthesis problem seeks to determine how to build a logical circuit from
basic logic gates (such as AND) that realizes a given Boolean function.
Given this logical circuit, we can obtain the corresponding tensor network
simply by replacing the gates with their tensor counterparts.
Using tensors that represent classical logical gates provides an alternative means to determine tensor networks representing for instance the GHZ and AND states \cite{CTNS}, compared to the MPS representations given in Examples \ref{ex:mps-ghz} and \ref{ex:w-state}.

\subsection[Penrose's tensor contraction algorithm]{Counting Graph Colorings} \label{sec:color}

Given a $3$-regular planar graph\footnote{A graph is $k$-regular iff every node has exactly $k$~edges connected to it.},
how many possible edge colorings using three colors exist,
such that all edges connected to each node have distinct colors?
This counting problem can be solved in an interesting (if not computationally efficient) way
using the order-3 $\epsilon$~tensor, which is defined in terms of components as
\begin{align}
\notag
 &\epsilon_{012} = \epsilon_{120} = \epsilon_{201} = 1,\\
 &\epsilon_{021} = \epsilon_{210} = \epsilon_{102} = -1,
\end{align}
otherwise zero. The counting algorithm is stated as
\begin{theorem}[Planar graph $3$-colorings, Penrose 1971 \cite{Penrose}]\label{thm:3-color}
The number~$K$ of proper $3$-edge-colorings of a planar $3$-regular graph
is obtained by replacing each node with an order-3 epsilon tensor,
replacing each edge with a wire,
and then contracting the resulting tensor network.
\end{theorem}

We will first consider the simplest case, a graph with just two nodes.
In this case we obtain
\be 
\includegraphics{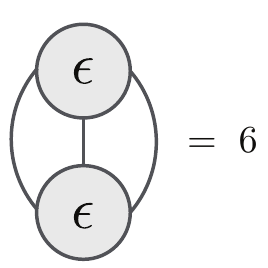}
\ee 
There are indeed $6$ distinct edge colorings for this graph, given as
\be
\includegraphics{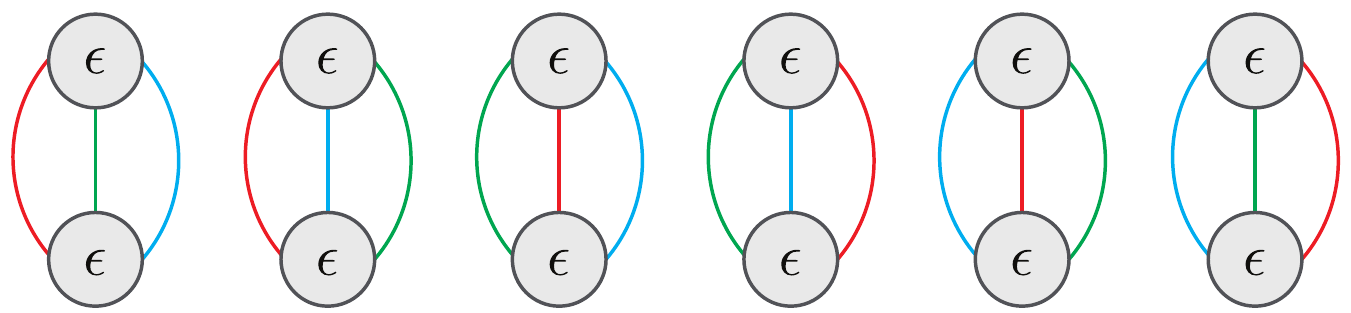} 
\ee

To understand Theorem~\ref{thm:3-color},
note first that the contraction~$K$ of the epsilon tensor network is the sum of all possible individual assignments of the index values
to the epsilon tensors comprising the network.
Each of the three possible index values can be understood as a color choice for the corresponding edge.
Whenever the index values for a given epsilon tensor are not all different, the corresponding term in~$K$ is zero.
Hence only allowed color assignments result in nonzero contributions to~$K$,
and for a graph that does not admit a proper $3$-edge-coloring we will have~$K=0$.
For instance, for the non-$3$-colorable Petersen graph we obtain
\be
\includegraphics{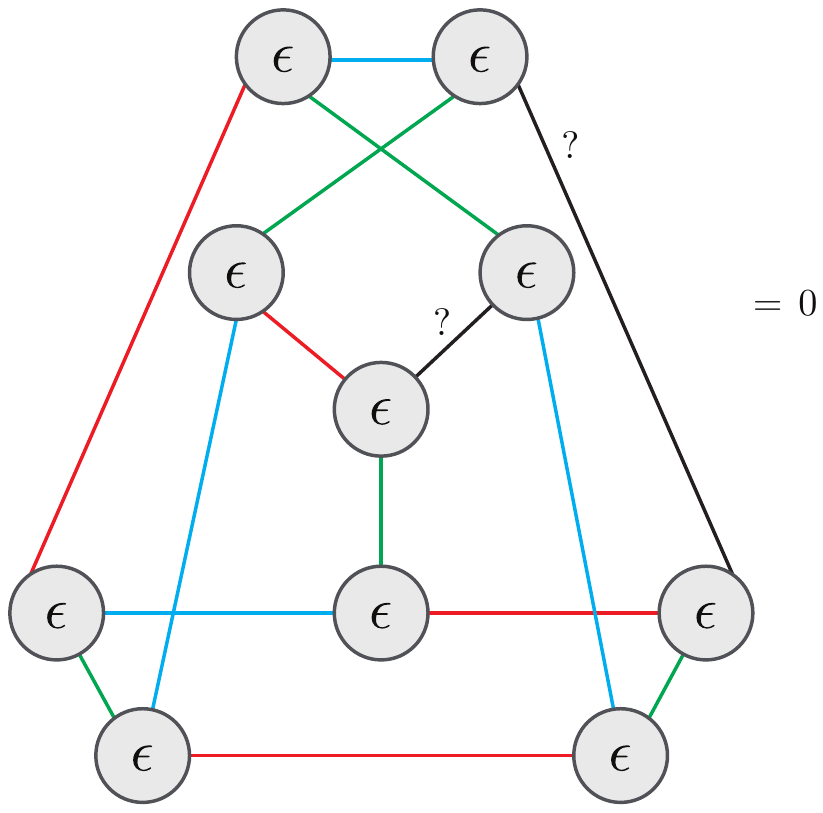}
\ee 

However, for~$K$ to actually equal the number of allowed colorings, each nonzero term must have the value~$1$ (and not~$-1$).
This is only guaranteed if the graph is planar,
as can be seen by considering the non-planar graph $K_{3,3}$:
\be
\includegraphics{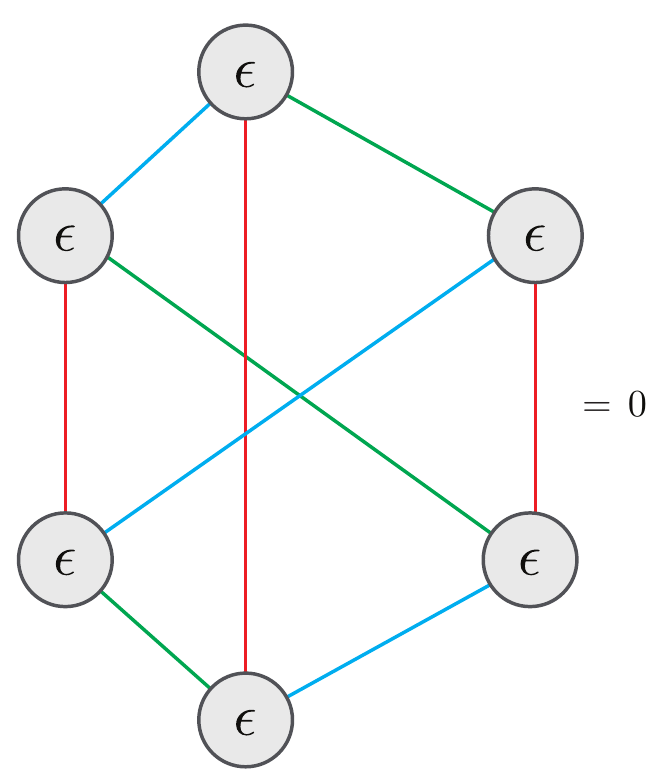}
\ee 
The edges can be colored with three colors---in 12 different ways---yet the contraction vanishes.

The computational complexity of this problem has been studied in \cite{xia2007computational}. Interesting, by a well known result (Heawood 1897), the 3-colorings as stated above, are one quarter of the ways of coloring the faces of the graph with four colors, so that no 
 two like-colored faces have an edge in common.  

\begin{example}[Physical implementation of $\epsilon_{abc}$ in quantum computing]
In quantum computing, typically one works with qubits (two level quantum systems) but implementations using qutrits exist (three level quantum systems, available in e.g.~nitrogen vacancy centers in diamond---see for instance \cite{2014NatCo...5E3371D}). 
The epsilon tensor $\epsilon_{abc}$ could be realized directly as a locally invariant 3-party state using qutrits, and can also be embedded into a qubit system.  We leave it to the reader to show that by pairing qubits, $\epsilon_{abc}$ can be represented with six qubits, where each leg now represents a qubit pair.  (Note that a basis of 3 states can be isometrically embedded in 4-dimensional space in any number of ways.) Show further that the construction can be done such that the two qubit pairs (together representing one leg) are symmetric under exchange.
Note the similarities with Eq.~\eqref{eqn:projaklt} in Example~\ref{ex:AKLT}.
\end{example}


\section{Frontiers in Tensor Networks}

Tensor network methods represent a vibrant area of active research, with new results and ideas appearing regularly. Here we have covered the elementary aspects of the tensor network language and three applications.  The first was the matrix product state representation, the next was tensor contractions to count Boolean formula solutions and the final application focused on tensor contractions to evaluate 3-edge-colorings 
of 3-regular planar graphs.
These three sample applications should provide a good base to move forward into active research.  
 
The most common tensor network structures and algorithms used in quantum mechanics include
Matrix Product States (MPS)~\cite{1992CMaPh.144..443F,1995PhRvL..75.3537O, 1997PhRvB..55.2164R} (see~Section~\ref{sec:dmps}) and the related
Density Matrix Renormalization Group (DMRG)~\cite{2011AnPhy.326...96S},
Matrix Product Operators (MPO)~\cite{2004PhRvL..93t7204V}, Tensor Trains~\cite{MAL-059}, 
Tree Tensor Networks (TTN)~\cite{2006PhRvA..74b2320S},
the Multiscale Entanglement Renormalization Ansatz (MERA)~\cite{Vidal2010,2007PhRvL..99v0405V,2008PhRvL.101r0503G,2008PhRvL.101k0501V},
Projected Entangled Pair States (PEPS)~\cite{2006PhRvL..96v0601V},
Correlator Product States (CPS)~\cite{2011PhRvB..84t5108A}
and Time-Evolving Block Decimation (TEBD)~\cite{2003PhRvL..91n7902V}---see
also time-evolution with MERA~\cite{2008PhRvA..77e2328R,2014GReGr..46.1823M}.   

Our reference list above is admittedly very much incomplete but should provide a solid starting place for further study.  Going further, there are several reviews that we encourage readers to consult.  These proceed largely towards approaches that map lattice problems to tensor networks as tools to solve models of strongly correlated systems \cite{2016arXiv160303039B, 2014AnPhy.349..117O, Vidal2010, MPSreview08, TNSreview09, 2011AnPhy.326...96S, Schollw, 2014EPJB...87..280O}---see also the viewpoint \cite{2010arXiv1006.0675S}.
There is a growing and active community exploring the use of tensor network algorithms as a means to discover and understand new properties of quantum systems.  

In terms of the graphical language, category theory is a branch of mathematics well suited to describe a wide range of networks \cite{2009arXiv0908.2469B}.  Quantum circuits were first given a `categorical model' in pioneering work by Lafont in 2003 \cite{Lafont03towardsan} and dagger compact closed categories \cite{BD95}, also called Baez-Dolan $\dagger$-categories, were first derived to describe both standard quantum theory as well as classes of topological quantum field theories in seminal work published in 1995 \cite{BD95}. (See \cite{2009arXiv0908.2469B} for a well written review of categorical quantum mechanics.)  For practical purposes, the graphical language turns out to be mathematically equivalent to the categorical formulation.

There has been some recent excitement surrounding
MERA~\cite{Vidal2010,2007PhRvL..99v0405V,2008PhRvL.101r0503G,2008PhRvL.101k0501V}---which is capable of representing the ground state of certain many-body models at their critical points---and its connection to quantum gravity research. Several interesting discoveries \cite{2012PhRvD..86f5007S} have recently been made around the so called tensor network incarnation of the AdS/MERA correspondence; networks which realize a discrete
anti-de Sitter space have a corresponding MERA network which represents the ground state of a critical system \cite{2012PhRvD..86f5007S, 2015PhRvD..91l5036B}. This has generated significant recent interest and excitement.  If this sounds exciting to you, you might want to take a look at \cite{2016arXiv160900026V,2014EPJB...87..280O, 2011JSP...145..891E}.

%
%

\begin{acknowledgments}
We've had a number of excellent collaborators over the years, who certainly influenced our understanding of the topic.  The least we can do is thank them here.
In alphabetical order we thank John Baez, Stephen Clark, Sam Denny,
Dieter Jaksch, Tomi Johnson, Marco Lanzagorta, Jason Morton, Lea Trenkwalder, Jacob Turner, Chris Wood and Zolt\'an Zimbor\'as,
as well as others we're probably forgetting. J.B.~acknowledges AFOSR grant FA9550-16-1-0300, Models and Protocols for Quantum Distributed Computation, for financial support.   Diagrams and cover courtesy of Lusa Zheglova (illustrator).  
\end{acknowledgments}

\bibliography{tensor-bib}

\clearpage
\appendix
\section{Tensors and Tensor Products}
\label{app:tensor}

The definition of a tensor starts with the \emph{tensor product}~$\otimes$.
There are many equivalent ways to define it, but perhaps the simplest one is
through basis vectors.
Let $V$ and~$W$ be finite-dimensional vector spaces over the same field of scalars~$\K$.
In physics-related applications~$\K$ is typically either the real numbers~$\R$
or the complex numbers~$\C$.
Now $V \otimes W$ is also a vector space over~$\K$.
If $V$ and~$W$ have the bases
$\{e_j\}_j$ and $\{f_k\}_k$, respectively,
the symbols $\{e_j \otimes f_k\}_{jk}$ form a basis for~$V \otimes W$.
Thus, for finite-dimensional spaces
$\dim(V \otimes W) = \dim V \, \dim W$.

The tensor product of two individual vectors $v \in V$ and $w \in W$
is denoted as $v \otimes w$.
For vectors the tensor product is a bilinear map
$V \times W \to V \otimes W$,
i.e.~one that is linear in both input variables.
For finite-dimensional spaces
one can obtain the standard basis coordinates of the tensor product of two vectors
as the \emph{Kronecker product} of the standard basis coordinates of the individual vectors:
\be
(v \otimes w)^{jk} = v^j w^k.
\ee
It is important to notice that due to the bilinearity $\otimes$ maps
many different pairs of vectors~$(v,w)$ to the same product vector:
$v \otimes (s w) = (s v) \otimes w = s (v \otimes w)$, where $s \in \K$.
For inner product spaces (such as the Hilbert spaces encountered in
quantum mechanics) the tensor product space inherits the inner product
from its constituent spaces:
\be
\inprod{v_1 \otimes w_1}{v_2 \otimes w_2}_{V \otimes W} = \inprod{v_1}{v_2}_V \inprod{w_1}{w_2}_W.
\ee

A \emph{tensor}~$T$ is an element of the tensor product of a finite
number of vector spaces over a common field of scalars~$\K$.
The dual space~$V^*$ of a vector space~$V$ is defined as the space of
linear maps from~$V$ to~$\K$. It is not hard to show that $V^*$~is a
vector space over~$\K$ on its own. This leads us to define
the concept of an order-$(p,q)$ tensor, an element of the tensor product
of $p$~primal spaces and $q$~dual spaces:
\be
T \in W_1 \otimes W_2 \otimes \ldots \otimes W_p \otimes V_1^* \otimes V_2^* \otimes \ldots \otimes V_q^*.
\ee
Given a basis $\{\bv\spidx{i}{}{k}\}_k$ for each vector
space~$W_i$ and a dual basis $\{\dv\spidx{i}{k}{}\}_k$ for each dual space~$V_i^*$,
we may expand~T in the tensor products of these basis vectors:
\be
\label{eq:tensorexpansion}
T = T\indices{^{i_1 \ldots i_p}_{j_1 \ldots j_q}}
\bv\spidx{1}{}{i_1} \otimes \ldots \otimes \bv\spidx{p}{}{i_p}
\otimes
\dv\spidx{1}{j_1}{} \otimes \ldots \otimes \dv\spidx{q}{j_q}{}.
\ee
$T\indices{^{i_1 \ldots i_p}_{j_1 \ldots j_q}}$ is simply an array of
scalars containing the basis expansion coefficients.
Here we have introduced the \emph{Einstein summation convention}, in which
any index that is repeated exactly twice in a term, once up, once
down, is summed over. This allows us to save a considerable number of
sum signs, without compromising on the readability of the formulas.
Traditionally basis vectors carry a lower (covariant) index and dual basis vectors
an upper (contravariant) index.

A tensor is said to be \emph{simple} if it can be written as the tensor product
of some elements of the underlying vector spaces:
$T = v\spidx{1}{}{} \otimes \ldots \otimes v\spidx{q}{}{} \otimes \varphi\spidx{1}{}{} \otimes \ldots \otimes \varphi\spidx{p}{}{}$.
This is not true for most tensors; indeed, in addition to the
bilinearity, this is one of the properties that separates tensors from mere Cartesian products of vectors.
However, any tensor can be written as a linear combination of simple tensors,
e.g.~as in Eq.~\eqref{eq:tensorexpansion}.

For every vector space~$W$ there is a unique bilinear map
$W \otimes W^* \to \K$,
$w \otimes \phi \mapsto \phi(w)$
called a natural pairing,
where the dual vector maps the primal vector to a scalar.
One can apply this map to any pair of matching primal and dual spaces in a tensor.
It is called a \emph{contraction} of the corresponding upper and lower indices.
For example, if we happen to have $W_1 = V_1$ we may contract the corresponding indices on~$T$:
\begin{align}
\notag
C_{1,1}(T)
&=
T\indices{^{i_1 \ldots i_p}_{j_1 \ldots j_q}}
\dv\spidx{1}{j_1}{}(\bv\spidx{1}{}{i_1})
\:\:
\bv\spidx{2}{}{i_2} \otimes \ldots \otimes \bv\spidx{p}{}{i_p}
\otimes
\dv\spidx{2}{j_2}{} \otimes \ldots \otimes \dv\spidx{q}{j_q}{}\\
&=
T\indices{^{k \, i_2 \ldots i_p}_{k \, i_2 \ldots j_q}}
\bv\spidx{2}{}{i_2} \otimes \ldots \otimes \bv\spidx{p}{}{i_p}
\otimes
\dv\spidx{2}{j_2}{} \otimes \ldots \otimes \dv\spidx{q}{j_q}{},
\end{align}
since the defining property of a dual basis is
$\dv\spidx{1}{j_1}{}(\bv\spidx{1}{}{i_1}) = \delta\indices{^{j_1}_{i_1}}$.
Hence the contraction eliminates the affected indices ($k$~is summed over),
lowering the tensor order by~$(1,1)$.

We can see that an order-$(1,0)$ tensor is simply a vector,
an order-$(0,1)$ tensor is a dual vector,
and can define an order-$(0,0)$ tensor to correspond to a plain scalar.
But what about general, order-$(p,q)$ tensors?
How should they be understood?
Using contraction, they can be immediately
reinterpreted as multilinear maps from vectors to vectors:
\begin{align}
\label{eq:Tmultilin}
\notag
T':\quad& V_1 \otimes \ldots \otimes V_q \to W_1 \otimes \ldots \otimes W_p,\\
T'(v\spidx{1}{}{} \otimes \ldots \otimes v\spidx{q}{}{}) &=
T\indices{^{i_1 \ldots i_p}_{j_1 \ldots j_q}}
\bv\spidx{1}{}{i_1} \otimes \ldots \otimes \bv\spidx{p}{}{i_p}
\times
\dv\spidx{1}{j_1}{}(v\spidx{1}{}{}) \times \ldots \times \dv\spidx{q}{j_q}{}(v\spidx{q}{}{}),
\end{align}
where we tensor-multiply $T$ and the vectors to be mapped together, and then contract the corresponding indices.
However, this is not the only possible interpretation.
We could just as easily see them as mapping dual vectors to dual vectors:
\begin{align}
\label{eq:Tmultilin2}
\notag
T'':\quad& W^*_1 \otimes \ldots \otimes W^*_p \to V^*_1 \otimes \ldots \otimes V^*_q,\\
T''(\varphi\spidx{1}{}{} \otimes \ldots \otimes \varphi\spidx{p}{}{}) &=
T\indices{^{i_1 \ldots i_p}_{j_1 \ldots j_q}}
\varphi\spidx{1}{}{}(\bv\spidx{1}{}{i_1}) \times \ldots \times \varphi\spidx{p}{}{}(\bv\spidx{p}{}{i_p})
\times
\dv\spidx{1}{j_1}{} \otimes \ldots \otimes \dv\spidx{q}{j_q}{}.
\end{align}
Essentially we may move any of the vector spaces to the
other side of the arrow by taking their dual:
\begin{align}
\label{eq:tensor-interp}
&W \otimes V^*
\quad \isom \quad
\K \to W \otimes V^*
\quad \isom \quad
V \to W
\quad \isom \quad
V \otimes W^* \to \K
\quad \isom \quad
W^* \to V^*,
\end{align}
where all the arrows denote linear maps.
Any and all input vectors are mapped to scalars by the corresponding
dual basis vectors in expansion~\eqref{eq:tensorexpansion}, whereas all input dual vectors map the
corresponding primal basis vectors to scalars.

If we expand the input vectors~$v\spidx{k}{}{}$ in
Eq.~\eqref{eq:Tmultilin} using the same bases as when expanding the tensor~T,
we obtain the following equation
for the expansion coefficients:
\begin{align}
T'(v\spidx{1}{}{} \otimes \ldots \otimes v\spidx{q}{}{})\indices{^{i_1 \ldots i_p}}
&=
T\indices{^{i_1 \ldots i_p}_{j_1 \ldots j_q}}
v\spidx{1}{j_1}{}
\cdots
v\spidx{q}{j_q}{}.
\end{align}
This is much less cumbersome than Eq.~\eqref{eq:Tmultilin}, and
contains the same information.
This leads us to adopt the \emph{abstract index notation} for tensors,
in which the indices no longer denote the components of the tensor in a
particular basis, but instead signify the tensor's order.
Tensor products are denoted by simply placing the tensor symbols next to each other.
Within each term, any repeated index symbol must appear once up and once down, and denotes
contraction over those indices.
Hence, $x^a$~denotes a vector (with one contravariant index),
$\omega_a$ a dual vector (with one covariant index),
and $T\indices{^{ab}_c}$ an order-$(2,1)$ tensor with two
contravariant and one covariant indices.
$S\indices{^{ab}_{cde}} x^c y^d P\indices{^e_a}$ denotes the contraction of
an order-$(2,3)$ tensor~$S$,
an order-$(1,1)$ tensor~$P$,
and two vectors, $x$ and $y$, resulting in an order-$(1,0)$ tensor
with one uncontracted index,~$b$.

In many applications, for example in differential geometry, the vector spaces
associated with a tensor are often copies of the same vector space~$V$
or its dual~$V^*$,
which means that any pair of upper and lower indices can be contracted,
and leads to the tensor components transforming
in a very specific way under basis changes.
This specific type of a tensor is called an
order-$(p,q)$ tensor \textit{on the vector space}~$V$.
However, here we adopt a more general definition, allowing $\{V_k\}_k$ and
$\{W_k\}_k$ to be all different vector spaces.

\end{document}